\documentclass[useAMS,usenatbib]{mn2e}

\usepackage{psfig, epsf, epsfig}

\title[Star formation in dense stellar systems]{Star formation 
within globular clusters: 
discrete multiple bursts and top-light 
mass functions}
\author[K. Bekki]
{Kenji Bekki${}^1$\thanks{E-mail:
kenji.bekki@uwa.edu.au} \\
${}^1$ICRAR M468
The University of Western Australia
35 Stirling Hwy, Crawley
Western Australia 6009, Australia}

\begin{document}

\date{Accepted, Received 2005 February 20; in original form }

\pagerange{\pageref{firstpage}--\pageref{lastpage}} \pubyear{2005}

\maketitle

\label{firstpage}

\begin{abstract}

The observed discrete multiple stellar populations and internal abundance
spreads
in $r$- and $s$-process elements within
globular clusters (GCs) have been suggested to be explained self-consistently
by discrete star formation events over a longer timescale ($\sim 10^8$ yr).
We here investigate whether such star formation is really possible
within GCs using numerical simulations
that include effects of dynamical interaction between
individual stars and the accumulated gas (``star-gas interaction'')
on star formation.
The principal results
are as follows. 
Small gas clouds 
with densities larger than $10^{10}$ atoms cm$^{-3}$ corresponding to
first stellar cores can be developed from gas without turbulence.
Consequently, new stars can be formed from the gas with
high star formation efficiencies ($>0.5$) in a bursty manner.
However, star formation can be suppressed  when
the gas mass fractions within GCs ($f_{\rm g}$) 
are less than a threshold value ($f_{\rm g, th}$).
This $f_{\rm g, th}$ is larger for GCs with lower masses and larger
gas disks.
Star-gas interaction and gravitational potentials
of GCs can combine to suppress the formation of massive
stars (i.e., ``top-light''
stellar initial mass function).
Formation of He-rich stars directly from
gas of massive AGB stars is possible in  massive GCs due to
low $f_{\rm g, th}$ ($<0.01$).
Short bursty star formation only for $f_{\rm g}>f_{\rm g, th}$
can be  partly responsible for discrete multiple star formation events within GCs.
We discuss how these results depend on the adopted model assumptions,
such as rotating gas disks within GCs.
\end{abstract}


\section{Introduction}

\begin{table*}
\centering
\begin{minipage}{180mm}
\caption{Possible positive 
and negative effects of existing (1G) stars within GCs on the formation of
later generations (LG; 2G, 3G etc) of  stars from gas accumulated within GCs.
}
\begin{tabular}{lll}
Physical process &  Possible effects  &  Reference \\
Global potentials of GCs  & Retention of  gas ejected from stars  &  D08, B11\\
Capture of cold gas  & Gas fueling for star formation  &  BM09, PK09 \\
Heating by compact objects  & Expulsion of cold gas from GCs   &  L13 \\
Radiative feedback effects of young 1G stars   &  Heating and evaporation of cold gas   & CS11 \\
Thermal and kinetic feedback of AGB stars    &  Removal of gas from GCs    &  D08, B11, B17b \\
Type Ia SN from 1G stars    &  Heating and expulsion of gas    &  D08, D16 \\
Type II SN from multiple generations of stars    &  Heating and expulsion of gas    &  BJK17, B18 \\
Delayed SNII   & Rapid removal of AGB ejecta & D16, DDV16 \\
Collisions of 1G stars with PMS stars  & Disruption of accretion disks  & T15 \\
Direct stellar bombardment (SB) on  gas clouds  & Tidal perturbation to growing gas clouds  &  This work \\
\end{tabular}
\end{minipage}
\end{table*}

It is one of crucial questions in observational and theoretical
studies of globular clusters (GCs)
why the Galactic GCs are observed to
show internal abundance spreads
in various elements to different degrees (Gratton et al. 2012 for a review).
The vast majority of GCs
clearly show internal abundance spreads
only in light elements (e.g., Caretta et al. 2009),
however,  some Galactic GCs have been observed
to show  such spreads
even in  [Fe/H] (e.g., Da Costa et al. 2009; Marino et al. 2009, 2011, 2018;
Johnson et al. 2015;  Yong et al. 2014).
These ``Fe-anomalous'' GCs also show internal abundance spreads
in $s$-process elements (e.g.,  Marino et al. 2011; Yong et al. 2014; 
Lardo et al. 2013), though Carretta et al. (2015) have found
internal spreads in $s$-process elements (e.g., La) in M80,
which is not classified as a Fe-anomalous GC.
Furthermore,
internal abundance spreads have been found in
helium for some GCs (e.g., Piotto et al. 2005; Milone et al. 2017)
and in $r$-process elements for 6 GCs (e.g., Snedin et al. 1997;
Roederer 2011).

It is one of goals for theoretical studies of GC formation to construct
a model that explains 
self-consistently
(i) the origin of the almost universal anti-correlations between light elements
(Na-O and Mg-Al)
and (ii) internal abundance spreads 
in $r$- and $s$-process elements and iron
observed in a significant fraction of the Galactic GCs.
One of popular scenarios for the formation of GCs with
multiple stellar populations (MSPs) is that
second generations  (``2G'') of stars were formed 
from gas ejected from first generations (``1G'') in the early formation histories
of GCs.  The chemical abundances of 2G stars
depend on  various factors, such as whether the gas from 1G stars
was mixed with pristine gas or not.
These self-enrichment processes have been now investigated by
many theoretical models of GC formation.

Various scenarios have been proposed so far to explain the observed
properties of GCs with MSPs (e.g., Renzini et al. 2015; Bastian \& Lardo 2018 for
recent reviews). Recent observational studies of internal abundance
spreads in $r$-process elements (e.g., Roederer 2011;
Sobeck et al. 2011; Worley et al. 2013) have been
suggested to provide very strong constrains on the theoretical models
of GCs:  formation scenarios in which self-enrichment can occur
before merging between neutron stars (earlier than $10^7$ yr after
1G star formation) could be possibly
ruled out (e.g., Bekki \& Tsujimoto 2017, BT17).
Furthermore, the observed discrete subpopulations of GCs
(e.g., Caretta 2014; Carretta et al. 2018) can be explained by discrete epochs
of star formation from AGB ejecta  mixed with pristine gas
(e.g., Bekki et al. 2017, BJP17).
Top-light initial mass functions (IMFs) of stars in intra-cluster medium (ICM)
with lower densities can be responsible for the prolonged star formation
within GCs in BJP17,
though such IMF variations are still being discussed 
extensively (e.g., Kroupa et al. 2013
for detailed discussion on this issue).
Kim \& Lee (2018) have recently proposed that star formation from
AGB ejecta mixed with stellar winds of massive stars and gas left
from original star-forming molecular clouds
is responsible for the origin of discrete MSPs within GCs.

D'Antona et al. (2016, D16) have  proposed a 
GC formation scenario in which  Fe-rich ejecta from SNIa or delayed SNII
can be mixed with AGB ejecta to form 2G (and 3G) stars with their Fe abundances
being slightly higher than those of 1G stars. In their scenario,
delayed SNII can truncate star formation from AGB ejecta: influences of
SNII from 2G are, however, completely ignored.
These recent works
have demonstrated that the AGB scenario can be the most promising candidate
that can explain anti-correlations between abundances of light elements,
discrete populations, and internal abundance spread in various elements 
(e.g., $s$-process) within GCs.
However, it is fair to say that it is yet to be determined which scenario
is the most realistic and reasonable one, mainly because the details of 
relevant physical processes in each scenario, in particular,
secondary star formation in gas from 1G stars,  have not been extensively
investigated.

One of possibly important physical processes related to star formation in GCs
is dynamical interaction between individual GC stars and collapsing star-forming
gaseous clouds within GCs. Owing to the very high number density of stars in a GC
($>10^4$ stars pc$^{-3}$), 
direct encounters of stars with gravitationally collapsing gas clouds
can influence the growth processes of such clouds: this situation is 
dramatically different from star formation in isolated fractal molecular
clouds with no existing stars, which have been investigated
by many authors (e.g., Klessen et al. 1998; Dale et al. 2011).
It would be possible that
such star-cloud interaction tidally disturbs the surrounding regions
of collapsing gas clouds and consequently suppress the growth of the clouds.
Cumulative effects of numerous  star-cloud interaction (referred to as ``stellar
bombardment'', SB) could have some dramatic effects on the evolution of gas accumulated
into GCs under some circumstances.
Previous numerical simulations could not investigate this possibly important
physical process, because they could not resolve the gas  density of $10^{10}$ atom
cm$^{-3}$ corresponding to the typical density of first stellar core
(e.g., Meyers 1978; Saigo et al. 2000)
owing to the relatively poor numerical resolutions:
B11 investigated secondary star formation on a scale of $1-10$ pc,
therefore it could not resolve the scale of first stellar  cores.

\begin{figure*}
\psfig{file=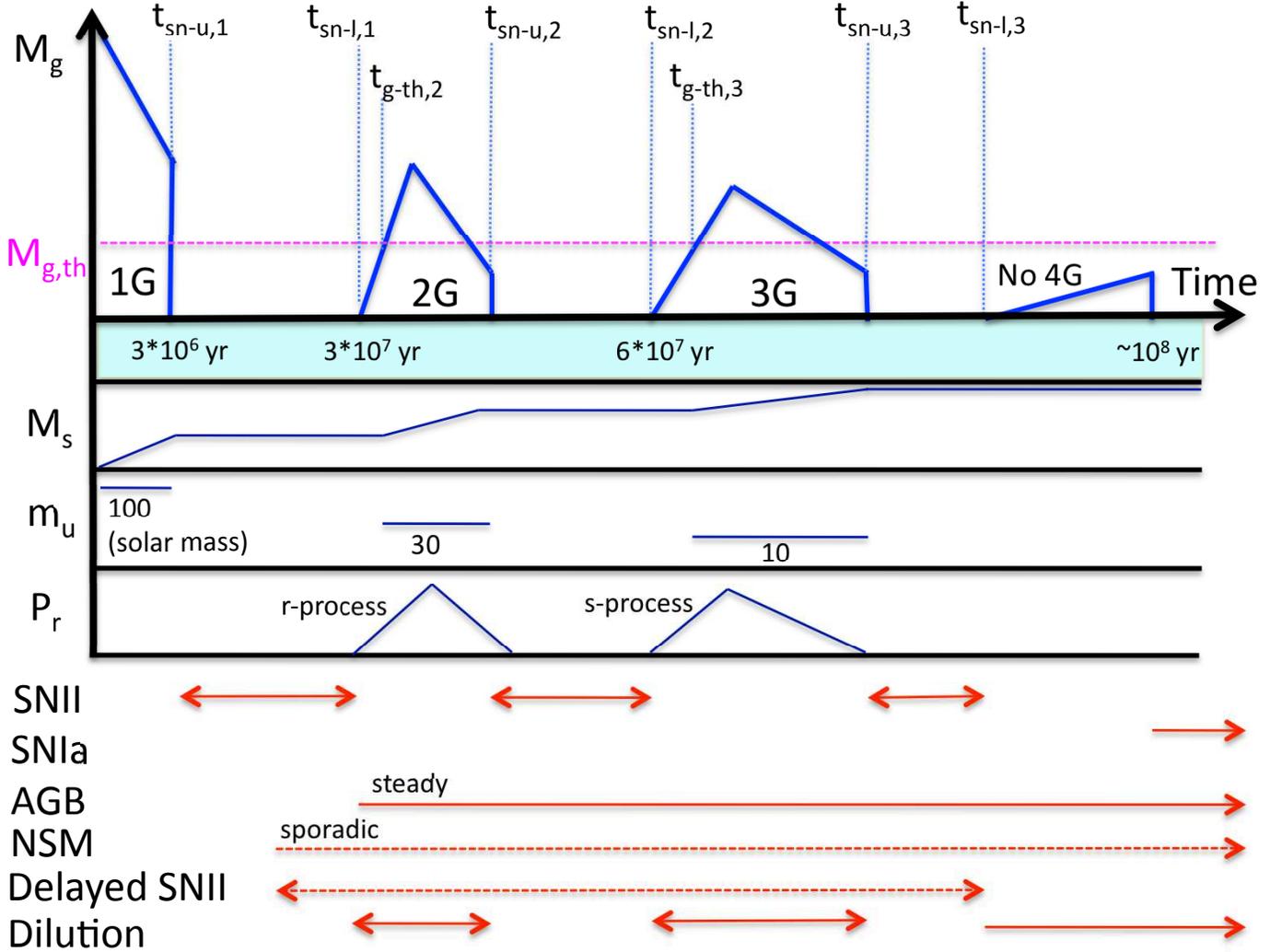,width=18.0cm}
\caption{
Illustration of the AGB scenario for the origins of discrete multiple
stellar populations (e.g., 1G means the first generation of stars)
and internal abundance spreads for
various elements (e.g., $s$- and $r$-process).
Time evolution of the total gas mass ($M_{\rm g}$), the 
total mass of new stars formed from gas 
($M_{\rm s}$), the upper-mass cut-off of the IMF ($m_{\rm u}$),
and the retention probability of gas that is consumed
by secondary star formation  ($P_{\rm r}$) is shown.
Red arrows indicate duration when each object (e.g., SNII) 
greatly influences and chemically pollutes ICM.
Gas ejection from AGB stars can steadily continue
due to contribution from the stars
with different masses
whereas gas ejection from neutron star merging (NSM) can occur sporadically.
Delayed core-collapse supernovae from binary massive stars
could occur sporadically too until 200 Myr after the formation of 1G stars
(e.g., Zapartas et al. 2017).
Multiple epochs
when dilution of AGB ejecta by pristine gas can be efficient
are also shown by red arrows.
The epochs of the most and least massive SNII from $i$-th  generation of 
stars are indicated by $t_{\rm sn-u, \it i}$ and 
$t_{\rm sn-l, \it i}$ respectively (here $i$=1, 2, and 3 only).
Also the time when $M_{\rm g}$ exceeds the threshold gas mass/fraction for star
formation ($M_{\rm g, th}$) is indicated for each discrete star formation
event (e.g., $t_{\rm g-th, 2}$ for 2G formation).
Star formation can be truncated either if SNe can expel the gas left
from star formation (for $m_{\rm u} \ge 8 {\rm M}_{\odot}$) or if
$M_{\rm g}$ or $f_{\rm g}$ (gas mass fraction) becomes less than
$M_{\rm g, th}$ or $f_{\rm g, th}$ 
(for $m_{\rm u} < 8 {\rm M}_{\odot}$).
A crucial parameter in this figure is the threshold gas mass ($M_{\rm g, th}$)
or mass fraction ($f_{\rm g, th}$) above which star  formation from
non-turbulent gas is possible within GCs.
Since gas ejection from AGB stars steadily continues, secondary
star formation from the gas is expected to continue steadily too.
However,  in this scenario, 
$M_{\rm g, th}$ and SNII does not allow such continuous star formation
under some circumstances (e.g., $m_{\rm u} \ge 8 {\rm M}_{\odot}$).
}
\label{Figure. 1}
\end{figure*}

The purpose of this paper is thus  to investigate, for the first time, 
how SB can influence the growth of gas clumps for 
star formation
in GCs. 
We particularly investigate whether and how gas clouds with densities
more than $10^{10}$ atoms cm$^{-3}$ corresponding to typical
densities of first stellar cores (e.g., Myers 1978; Saigo et al. 2001)
using our original hydrodynamical simulations of gas evolution
in such dense systems.
We also investigate how the results depend on (i) the original masses and
sizes of GCs and (ii) the gas mass fractions ($f_{\rm g}$) in GCs
in order to understand under what physical conditions secondary star formation
is possible.
This investigation can reveal the details of secondary star formation
processes and thus provide valuable insights for the origin of
MSPs in GCs.
Since this is the very first investigation on this important issue,
we do not include chemical evolution of gas during secondary star formation
within GCs. We also adopt a dynamical equilibrium model for all GC simulations,
which means that more realistic growth processes of cluster formation through
hierarchical merging of subclusters or cluster complex (e.g., Adamo et al. 2012;
Bekki 2017a, B17a) are totally ignored.

The SB effects on star formation can be only one  of several key effects
of stars on star formation within GCs.
Positive and negative effects of existing stars of GCs
on secondary star formation
have been discussed in previous studies and 
are briefly summarized in Table 1 for comparison.
These include (1) retention of gas due to global gravitational
potentials of GCs (e.g., D'Ercole et al. 2008, D08; Bekki 2011, B11),
(2) gravitational capture of cold gas by GCs 
(e.g., Bekki \& Mackey 2009, BM09; Pflamm-Altenburg \& Kroupa 2009, PK09,
Armstrong et al. 2018),
(3) heating of ICM by compact objects such as stellar mass black holes
(e.g., Leigh et al. 2013, L13), (4) radiative feedback effects of
young stars (Conroy \& Spergel 2010, CS11),
(5) feedback effects of AGB stars (D08; B11; Bekki 2017b,  B17b),
(6) SNIa explosions (D08, D16), 
(7), SNII explosions
from multiple generations of stars (BJK17; Balin 2018, B18), 
(8) disruption of accretion disks around pre-main-sequence stars (PMS)
by collisions of 1G stars (Tailo et al. 2015, T15),
and (9) delayed SNII (D16, D'Ercole et al., 2016, DDV16).

For \& Bekki (2017) have recently discovered young stellar 
objects  (YSOs) in younger star clusters in the LMC, which is evidence
that gas exists within the clusters (or it existed until quite recently).
This new observation can justify the adopted assumption of cold gas
within star clusters,
though no observations have ever revealed
H$\alpha$ emission from massive OB stars
within  star clusters in star-forming 
galaxies.
The observed lack of OB stars in young clusters is not necessarily
inconsistent with secondary star formation, because 
the upper-mass cut-off ($m_{\rm u}$) of the IMF
can be less than $30 {\rm M}_{\odot}$ during star formation within GCs
(BJK17).

The plan of the paper is as follows.
We describe the models for initial 3D distributions of gas from 1G
stars and for star formation from dense gas clouds
in \S 2.
We present the results of numerical simulations of
gas clump formation in GCs
in \S 3.
Based on these results,
we provide several implications of the present results
in the context of MSPs within GCs
in \S 4.
We summarize our  conclusions in \S 5.
In this paper, we focus exclusively on 
star formation from gas that is either from external gas accretion
or from AGB stars. We do not discuss star formation from gas within
GC-hosting molecular clouds
that are chemically polluted by  stellar winds of massive stars, though
our previous simulations demonstrated that
such star formation can be important for the abundance spreads
in He and CNO within GCs (e.g., Bekki \& Chiba 2007).

\begin{table}
\centering
\begin{minipage}{80mm}
\caption{The parameter values for a GC in
a representative models without star formation.
}
\begin{tabular}{ll}
Stellar mass  & $ 3.1 \times 10^5 {\rm M}_{\odot}$  \\
Scale length of the Plummer model   & $ 2 $ pc   \\
Lower-mass cut-off ($m_{\rm l}$) for  stars & $0.1 {\rm M}_{\odot}$ \\
Upper-mass cut-off for stars ($m_{\rm u}$) & $120 {\rm M}_{\odot}$ \\
Gas mass  & $3000 {\rm M}_{\odot}$  \\
Gas disk size  & 1 pc  \\
Mass resolution (gas)  &  $0.01 {\rm M}_{\odot}$ \\
Spatial resolution  &  0.014 pc \\
SB effect   &  Yes \\
SN and AGB feedback from 2G stars   & No \\
\end{tabular}
\end{minipage}
\end{table}

\begin{table}[h]
\centering
\begin{minipage}{85mm}
\caption{Description of the model parameters for  
20 models investigated in the present study.
}
\begin{tabular}{llll}
Model ID & $M_{\rm gc}$ (${\rm M}_{\odot}$) 
& $M_{\rm g}$ ($M_{\rm g}$) & comments \\ 
MN1 & $3.1 \times 10^5$ & $1.0 \times 10^3$   &  \\
MN2 &  $3.1 \times 10^5$   & $3.0 \times 10^3$  & fiducial  \\
MN3 &  $3.1 \times 10^5$   & $5.0 \times 10^3$  &   \\
MN4 & $3.1 \times 10^5$ & $1.0 \times 10^3$   &  W/O SB\\
MN5 &  $3.1 \times 10^5$   & $3.0 \times 10^3$  & W/O SB  \\
MN6 &  $3.1 \times 10^5$   & $5.0 \times 10^3$  & W/O SB  \\
MN7 & $-$ & $1.0 \times 10^3$   &  No 1G stars \\
MN8 &  $-$   & $3.0 \times 10^3$  & No 1G stars \\
MN9 &  $-$   & $5.0 \times 10^3$  & No 1G stars  \\
M1 & $3.1 \times 10^5$ & $1.0 \times 10^2$   &  \\
M2 &  $3.1 \times 10^5$   & $3.0 \times 10^2$  &   \\
M3 &  $3.1 \times 10^5$   & $1.0 \times 10^3$  &   \\
M4 &  $3.1 \times 10^5$   & $3.0 \times 10^3$  &   \\
M5 &  $3.1 \times 10^5$   & $5.0 \times 10^3$  &   \\
M6 &  $3.1 \times 10^5$   & $1.0 \times 10^3$  & $R_{\rm gc}=6.9$ pc  \\
M7 &  $1.5 \times 10^5$   & $1.0 \times 10^2$  &   \\
M8 &  $1.5 \times 10^5$   & $3.0 \times 10^2$  &   \\
M9 &  $1.5 \times 10^5$   & $1.0 \times 10^3$  &   \\
M10 &  $1.5 \times 10^5$   & $3.0 \times 10^3$  &   \\
M11 &  $1.5 \times 10^5$   & $5.0 \times 10^3$  &   \\
M12 &  $1.5 \times 10^5$   & $1.0 \times 10^3$  & $R_{\rm gc}=20$ pc  \\
M13 &  $3.1 \times 10^4$   & $1.0 \times 10^2$  &  \\
M14 &  $3.1 \times 10^4$   & $3.0 \times 10^2$  &   \\
M15 &  $3.1 \times 10^4$   & $1.0 \times 10^3$  &   \\
M16 &  $3.1 \times 10^4$   & $3.0 \times 10^3$  &   \\
M17 &  $3.1 \times 10^4$   & $1.0 \times 10^2$  & $R_{\rm gc}=4.6$ pc  \\
M18 &  $3.1 \times 10^4$   & $1.0 \times 10^2$  & $R_{\rm gc}=6.9$ pc  \\
M19 &  $3.1 \times 10^3$   & $1.0 \times 10^3$  &  \\
M20 &  $3.1 \times 10^2$   & $1.0 \times 10^3$  &   \\
M21 &  $8.1 \times 10^5$   & $3.0 \times 10^3$  &   \\
M22 &  $8.1 \times 10^5$   & $5.0 \times 10^3$  &   \\
M23 &  $8.1 \times 10^5$   & $1.0 \times 10^4$  &   \\
M24 &  $8.1 \times 10^5$   & $5.0 \times 10^4$  &   \\
M25 &  $1.6 \times 10^6$   & $1.0 \times 10^3$  &   \\
M26 &  $2.3 \times 10^6$   & $3.0 \times 10^3$  &   \\
M27 &  $2.3 \times 10^6$   & $5.0 \times 10^3$  &   \\
M28 &  $2.3 \times 10^6$   & $1.0 \times 10^4$  &   \\
M29 &  $2.3 \times 10^6$   & $5.0 \times 10^4$  &   \\
M30 &  $3.1 \times 10^5$   & $1.0 \times 10^3$  &   \\
M31 &  $3.1 \times 10^5$   & $3.0 \times 10^2$  &  W/O SB \\
M32 &  $3.1 \times 10^5$   & $1.0 \times 10^3$  &  W/O SB \\
M33 &  $3.1 \times 10^5$   & $3.0 \times 10^3$  &  W/O SB \\
M34 &  $3.1 \times 10^4$   & $1.0 \times 10^3$  &  W/O SB \\
M35 &  $3.1 \times 10^5$   & $3.0 \times 10^3$  &  $R_{\rm g}=2$ pc \\
M36 &  $3.1 \times 10^5$   & $3.0 \times 10^3$  &  $R_{\rm g}=3$ pc \\
M37 &  $3.1 \times 10^5$   & $1.0 \times 10^4$  &  $R_{\rm g}=3$ pc \\
M38 &  $3.1 \times 10^5$   & $4.0 \times 10^4$  &  $R_{\rm g}=3$ pc \\
M39 &  $8.1 \times 10^5$   & $3.0 \times 10^3$  &  $R_{\rm g}=3$ pc \\
M40 &  $8.1 \times 10^5$   & $3.0 \times 10^4$  &  $R_{\rm g}=3$ pc \\
M41 &  $8.1 \times 10^5$   & $8.0 \times 10^4$  &  $R_{\rm g}=3$ pc \\
M42 &  $2.3 \times 10^6$   & $3.0 \times 10^3$  &  $R_{\rm g}=3$ pc \\
M43 &  $2.3 \times 10^6$   & $3.0 \times 10^4$  &  $R_{\rm g}=3$ pc \\
M44 &  $2.3 \times 10^6$   & $1.0 \times 10^5$  &  $R_{\rm g}=3$ pc \\
\end{tabular}
\end{minipage}
\end{table}

\section{The model}

\subsection{The AGB scenario}
\subsubsection{Gas accumulation}
We adopt the ``AGB scenario'' in which AGB ejecta
from 1G  can be converted into
new stars to become 2G (e.g., D'Antona et al. 2002; Bekki et al. 2007, B07:
D08).
In this scenario,  AGB ejecta from 1G stars within GCs can be mixed with
pristine gas (with the chemical abundances being the same as those
of 1G stars), though such mixing is not necessary for some GCs.
The total gas mass ($M_{\rm g}$) within a GC is accordingly as follows:
\begin{equation}
M_{\rm g} = M_{\rm acc}+M_{\rm agb} \;
\end{equation}
where $M_{\rm acc}$ is the total mass of pristine gas
that have the same chemical abundances as those of 
the GC's host giant molecular cloud (GMC) and is accreted on the 
GC's central region  and $M_{\rm agb}$
is that of gas ejected from AGB stars.
$M_{\rm acc}$ is further divided into (i) the total mass of
gas that is left from 1G star formation within the GMC
(``internal'') and (ii) that of gas that is initially outside the GC
(``external'', e.g.,  ISM) 
as follows:
\begin{equation}
M_{\rm acc} = M_{\rm acc, int}+M_{\rm acc, ext} .
\end{equation}
Although B17b has shown that most of the remaining pristine gas within GC-hosting GMCs
can be expelled by energetic multiple SNII,
it could be possible that a minor fraction of the remaining gas 
can be retained within very massive GC-hosting GMCs.

The total mass of AGB ejecta  is divided into (i) the total mass of gas from
AGB stars within the GC (``internal'') and (ii) that of AGB stars that do never belong
to the GC (``external'') as follows:
\begin{equation}
M_{\rm agb} = M_{\rm agb, int}+M_{\rm agb, ext} .
\end{equation}
If GCs are formed in the central regions of dwarf galaxies embedded
in massive dark matter halos,
then gas ejected from field AGB stars within the dwarfs can be 
accumulated within GCs owing to deep gravitational
potentials of the dwarfs (e.g., Bekki 2006; Maxwell et al. 2014).
Our recent numerical simulations have also shown that about 30\% of $M_{\rm agb}$
in GCs
can originate from field AGB stars that never become the GC member stars
(Bekki 2018). 
Furthermore, if a GC is formed from a group of numerous
star clusters (i.e., hierarchical star cluster complex),
then AGB ejecta from some small clusters that do not finally belong to the GC 
can be also accreted onto the GC (B17a).
Thus, field AGB stars in GC-hosting galaxies can contribute significantly
to $M_{\rm agb, ext}$.
 In the present study,
we do not specify the relative contribution of these $M_{\rm agb, int}$
and $M_{\rm agb, ext}$.

The most important parameter in the present study is the mass fraction
of gas within GCs, which is as follows:
\begin{equation}
f_{\rm g} = M_{\rm g}/M_{\rm gc}  ,
\end{equation}
where $M_{\rm gc}$ is the total stellar mass of a GC.
If $M_{\rm g} = M_{\rm agb, int}$, then $f_{\rm g}$ is 
time-dependent and a function
a number of parameters, such as the stellar initial mass function (IMF)
and the gas accretion timescales within GCs (Fig. 1 in B11). 
This $f_{\rm g}$ can be as large as 0.05 for the Salpeter IMF $\sim 10^8$ yr
after the formation of 1G stars (B11). 
As discussed by D08 and D16, prompt SNIa can start to expel all of the 
gas within GCs $\sim 10^8$ yr after 1G formation. Therefore, the
maximum possible $f_{\rm g}$ could be $\sim 0.05$,
if $M_{\rm g}=M_{\rm agb, int}$.

\begin{figure*}
\psfig{file=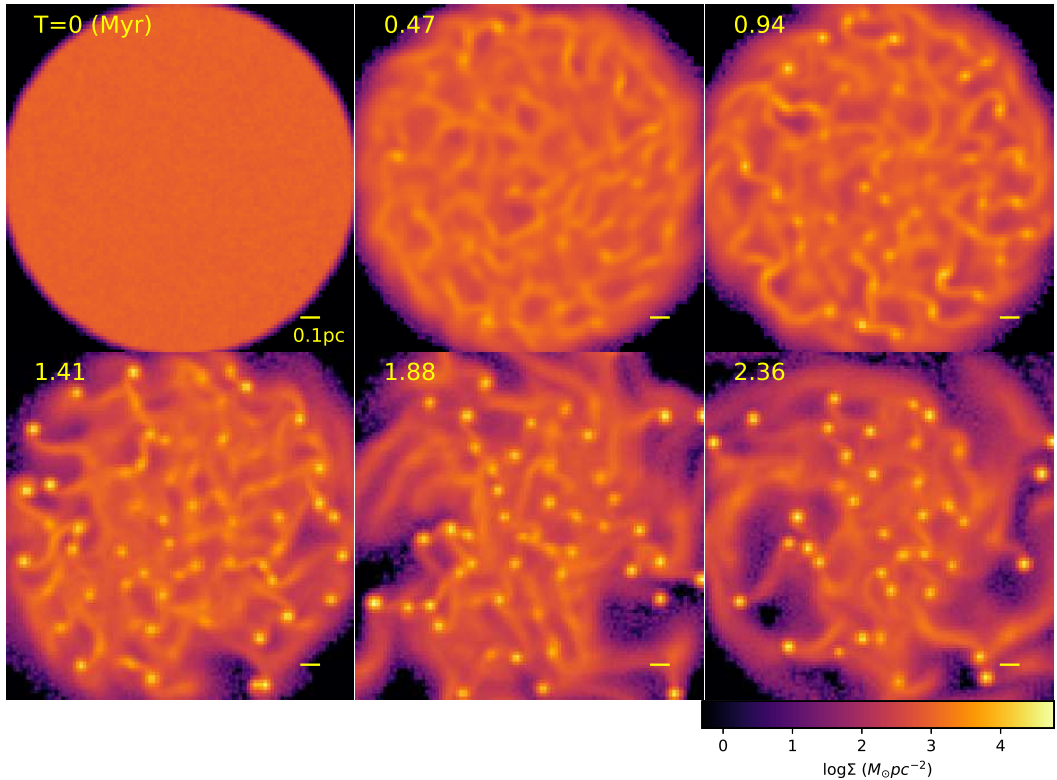,width=18.0cm}
\caption{
Time evolution of the surface mass density ($\Sigma$ in logarithmic scale)  of gas 
projected onto the $x$-$y$ plane for
the fiducial model MN2 without star formation.
The total gas mass within the GC is $3 \times 10^3 {\rm M}_{\odot}$,
which means that the gas mass fraction ($f_{\rm g}$) is 0.01.
Time ($T$) that has elapsed since the start of this simulation
is shown in the upper left corner of each frame.
The scale bar of 0.1 pc is shown in the lower right at each panel.
}
\label{Figure. 2}
\end{figure*}

\begin{figure*}
\psfig{file=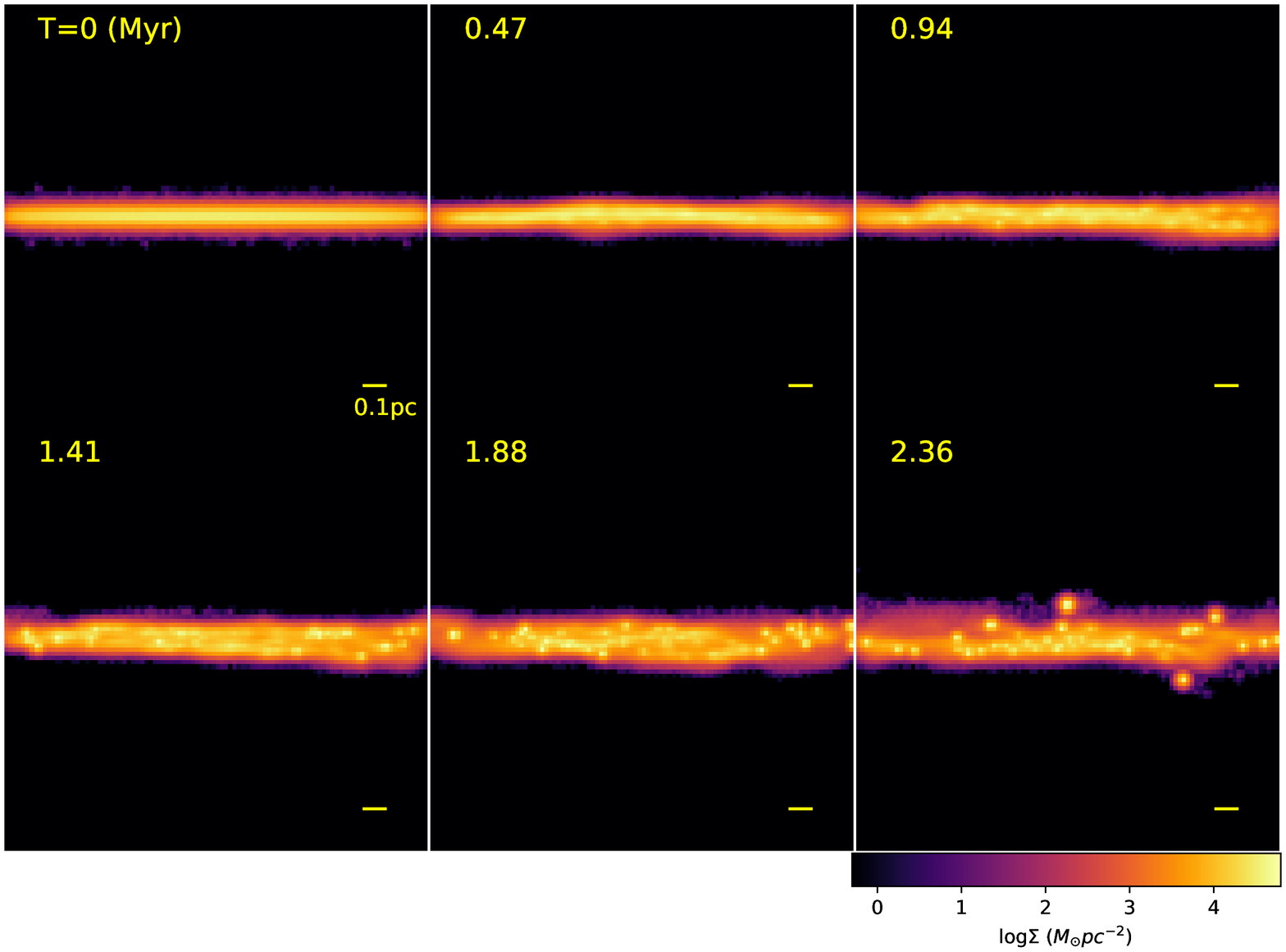,width=18.0cm}
\caption{
The same as Fig.2 but for the $x$-$z$ projection.
}
\label{Figure. 3}
\end{figure*}

\subsubsection{Time sequence of star formation}

Fig. 1 illustrates the time sequence of discrete multiple star formation within GCs
in the AGB scenario.
This is only for the case of GCs which form three generations of
stars (1G, 2G, and 3G): it should be noted here that some GCs have only
two or four generations depending on the gas accretion histories
of the GCs.
As discussed in detail later in this paper, 
a threshold $f_{\rm g}$ ($f_{\rm g, th}$) or $M_{\rm g}$ ($M_{\rm g, th}$)
above which star formation is possible
is the key parameter that determines
the star formation history of a GC.
Also, as discussed in BJP17, the upper-mass cut-off of the IMF 
($m_{\rm u}$) during each episode
of star formation can control the duration of 
the episode. The time sequence of star formation in the AGB scenario  is described
as follows (The starting time, $T=0$, corresponds to the onset of star formation
within GMC below). First, the formation
of 1G stars starts within  a fractal giant molecular cloud (GMC) and 
continues
until the explosion of the most massive stars ($m_{\rm s} \sim 100 {\rm M}_{\odot}$).
All of the remaining gas left from 1G formation is expelled from the forming GC
due to the energetic SN feedback effect: this epoch is defined as
$t_{\rm sn-u, 1}$.

Gas ejected from the most massive AGB stars
can start to be accreted onto the central region of the GC
after the least massive
SN ($m_{\rm s}=8 {\rm M}_{\odot}$) explodes ($T=t_{\rm sn-l, 1}$).
The gas mass steadily increase and exceeds  $M_{\rm g, th}$ at some point
so that the formation of 2G stars can start ($T=t_{\rm g-th, 2}$).
Therefore, there should be a time lag between the commencement of gas accretion
from AGB stars and that of active star formation from the 
accreted gas within GCs.
This time lag could be longer than the timescale of star formation
from the accreted gas within GCs.

This period of no star formation is denoted
as ``fallow period'' ($\Delta t_{\rm f,1}$) 
in the  AGB scenario, 
and the first fallow period  is  as follows:
\begin{equation}
\Delta t_{\rm f,1} = t_{\rm g-th, 2}-t_{\rm sn-u, 1}.
\end{equation}
The lifetime of the most massive star in each star formation
episode can be different due to different $m_{\rm u}$.
Also, the timescale of $M_{\rm g}$ to exceed $M_{\rm g, th}$
can be different in different star formation episodes.
Therefore, $\Delta t_{\rm f, i}$, where $i$ represents $i$-th generation
of stars, can be different between different star formation episodes.

The formation of 2G stars is truncated by energetic feedback
effects of the most massive SNII of 2G , and the duration
of 2G formation ($\Delta t_{\rm sf, 2}$) is given as follows.
\begin{equation}
\Delta t_{\rm sf,2} = t_{\rm sn-u, 2}-t_{\rm g-th, 2}.
\end{equation}
Star formation from accreted gas can continue as long
as $M_{\rm g}$ exceeds $M_{\rm g, th}$. 
This $\Delta t_{\rm sf, 2}$ depends strongly on $m_{\rm u}$,
which determines the lifetime of the most massive star.
For the $i$-th generation of stars, $t_{\rm sf, \it i}$ is as follows:
\begin{equation}
\Delta t_{\rm sf,\it i} = t_{\rm sn-u, \it i}-t_{\rm g-th, \it i}.
\end{equation}
In Fig. 1,
the formation of 4G stars is not possible,
because $M_{\rm g}$ cannot become larger than $M_{\rm g, th}$,
before the gas is expelled by SNIa.
The cessation of star formation by SNIa has been already
suggested in the earlier models based on the AGB scenario
(D08, D16).
Star formation from AGB ejecta
can be also truncated by delayed SNII (D16, DDV16),
because recent theoretical works have shown that about 15\% of core-collapse
supernovae occur $50-200$ Myr after star formation (e.g., Zapartas et a. 2017).

\subsubsection{Probability of gas retention}

In the AGB scenario,  gas ejected from AGB stars and NSM needs to be
retained for a sufficient time  so that the gas (mixed with pristine gas)
can be converted into new stars.
Such star formation can explain the observed 
abundance spreads in $r$- and $s$-process elements of GCs.
The probability of gas retention within GCs 
($P_{\rm r}$) depends on a number of
physical parameters, such as gravitational potential wells of GCs,
mass densities of ICM ($\rho_{\rm icm}$),
and ejection velocities of gas from stars.
As shown in BT17, $P_{\rm r}$ for $r$-process elements from NSM can be
high if $\rho_{\rm icm}$ is as large as $10^5$ atom cm$^{-3}$:
it depends on $\rho_{\rm icm}$ whether gas 
from NSM can
be retained within GCs (BT17).
The delay time distribution of NSM demonstrates that NSM can start to occur
around $\sim 10^7$ yr after star formation 
(e.g., Fig.14 in Dominik et al. 2012).
This means that gas from NSM of 1G needs to be mixed with high-density gas
either from AGB stars of 1G or from ISM (i.e., no original gas can be left
at the time of NSM).

Gas ejected from AGB stars with low 
wind velocities of $\sim 10$ km s$^{-1}$ 
can be retained within GCs with deeper gravitational potentials.
A larger amount of NSM ejecta can be retained in GCs with
higher $\rho_{\rm icm}$ (BT17). 
Therefore, $P_{\rm r}$ for AGB and NSM ejecta
is a function of $\rho_{\rm icm}$
and $\phi_{\rm gc}$) as follows.
\begin{equation}
P_{\rm r} = f(\rho_{\rm icm}, \phi_{\rm gc}),
\end{equation}
where the functional form $f$ can be investigated by numerical simulations
of interaction between gas ejected from NSM and AGB stars and ICM within GCs.
In order to explain the observed abundance patters of anomalous GCs with
[Fe/H] spreads,
D16 considered that pristine gas mixed with ejecta from SNIa or delayed SNII
can be converted into 2G stars with [Fe/H] slightly higher than that of 1G.
Such self-enrichment due to SNe within GCs appears to be highly unlikely,
given the large amount of energy from SNe ($10^{51}$ erg per SN)
and  the relatively shallow gravitational potential
wells of GCs.
However, if $\rho_{\rm icm}$ is rather high, such SN ejecta could interact with
ICM and consequently lose kinematic energy and momentum to finally be reaccreted onto
GCs.

\subsubsection{$m_{\rm u}$ as a key parameter}

If the IMF in 2G (or 3G, 4G etc) is a canonical one with the slope of $-2.35$ and
$m_{\rm u} \sim 100 {\rm M}_{\odot}$, then the duration of star formation
can be very short ($<3$ Myr) due to the truncation of 
star formation by energetic  feedback effects
of the massive SNe.
Therefore, it is possible that only a small fraction of $M_{\rm g}$ can be
converted into new stars within such a short timescale.
This possibly low SFE ($\epsilon_{\rm sf}$) exacerbates the mass budget problem
of the AGB scenario: even for $\epsilon_{\rm sf}$ is assumed to be 1,
the original GC mass (the mass of 1G stars) is by a factor of $\sim 10$
larger than the present-day GC mass to explain the large fraction of 2G stars
for a canonical IMF. 
Furthermore, $m_{\rm u}$ cam determine $\Delta t_{\rm sf, \rm i}$ ($i$=1, 2 etc),
because the time lag between star formation and SN explosion of the most 
massive stars  depends on $m_{\rm u}$.
It is possible that if $m_{\rm u}$ is less than $8 {\rm M}_{\odot}$,
star formation can continue within GCs for a quite long time 
until energetic events other than SNII (e.g., SNIa or delayed SNII).
Thus, $m_{\rm u}$ is a key parameter for star formation
of later generations of stars (2G, 3G etc) in the AGB scenario.

It could be possible that the slope of the IMF in later generations
(LG) of stars  (2G, 3G etc) could be different from that of 1G
owing to the formation of LG stars in dense stellar systems.
If the IMF in LG (2G, 3G etc) is different from that of 1G,
then the mass function (MF) of stars 
can be still different between 1G and LG populations even after 
$10$ Gyr dynamical evolution.
Vesperini et al. (2018) have recently investigated the evolution
of the MF  separately both for 1G and 2G stars within GCs using
numerical simulations.  They found that the initial difference of MFs
between 1G and 2G can be still visible in the present-day GCs,
though they evolve with time during  dynamical evolution of the GCs.
Such possible differences are yet to be detected observationally.

\subsection{Simulation code}
Based on the AGB scenario,
We investigate star formation from dense gas clouds in the central regions
of dense stellar systems (DSSs) such as GCs and stellar galactic nuclei
 using our original chemodynamical simulations codes that can be run
on GPU clusters (Bekki 2013, B13; Bekki 2015, B15). Since our main focus is not the evolution
of metals and dust in galaxies and star clusters, we ``switch off'' the 
components of the code that are relevant to evolution of metals and dust.
The code 
combines the method of smoothed particle
hydrodynamics (SPH) with calculations of three-dimensional
self-gravitating fluids in astrophysics.
Since the details of the code are given in B13, we just briefly describe
it in this paper.

It would be the best to investigate
both (i)  the formation of stars from fractal molecular clouds
(i.e, 1G)
and (ii)  the formation of new stars 
(2G) from gas ejected from stars or from interstellar medium (ISM)
in a self-consistent manner.
However, our recent simulations (B17a, b)
which have done this type
of self-consistent simulations, do not have enough mass and size
resolutions to investigate direct gravitational interaction between
individual stars and small gas clouds.
Therefore, we have adopted the present models in which GCs are initially
in dynamical equilibrium in order to grasp some essential ingredients
of such effects of star-gas interaction on secondary star formation
within GCs.

\subsection{Original stellar systems}

We assume that an original dense stellar system 
has a Plummer density profile
(e.g., Binney \& Tremaine 1987) with a mass ($M_{\rm gc}$),
and a size ($R_{\rm gc}$), and a central 
velocity dispersions (${\sigma}_{\rm gc}$).
The scale length ($a_{\rm gc}$) of the system is determined by the formula
\begin{equation}
a_{\rm gc} = GM_{\rm gc}/6{{\sigma}_{\rm gc}}^{2}, \;
\end{equation}
where G is the gravitational constant.
Observations showed that GCs have no strong relation between $M_{\rm gc}$ and
$R_{\rm gc}$
with  a large dispersion of $R_{\rm gc}$ for a given $M_{\rm gc}$
(e.g., Djorgovski et al.  1997;  Ashman \& Zepf  2001).
This no strong size-mass relation is suggested to be understood
in the context of different
star formation efficiencies in GCs with different $M_{\rm gc}$.
Therefore,
we mainly investigate the models with a fixed  $R_{\rm gc}$ that is
consistent with the observed typical GC size
for different $M_{\rm gc}$. However,
we also investigate  models with  different $R_{\rm gc}$ for a 
fixed $M_{\rm gc}$  in
the present study.

The stellar system is composed of stars with different masses
(not equal-mass, as assumed in B11) that follow the canonical
Salpeter initial mass function (IMF) of stars.
The adopted IMF in number is thus defined
as follows:
\begin{equation}
\psi (m_{\rm I}) = M_{gc,0}{m_{\rm I}}^{-\alpha},
\end{equation}
where $m_{\rm I}$ is the initial mass of
each individual star and the slope $\alpha =2.35$.
The normalization factor $M_{gc,0}$ is a function of $M_{\rm gc}$,
$m_{\rm l}$ (lower mass cut-off), and $m_{\rm u}$ (upper mass cut-off):
\begin{equation}
M_{gc,0}=\frac{M_{\rm gc}
\times (2-\alpha)}{{m_{\rm u}}^{2-\alpha}-{m_{\rm l}}^{2-\alpha}}.
\end{equation}
We mainly investigate the models in which
$m_{\rm l}$ and $m_{\rm u}$ are  set to be   $0.1 {\rm M}_{\odot}$
and  $120 {\rm M}_{\odot}$, respectively.
In the models with
 these IMF mass cut-offs, $M_{\rm gc}$ is $3.1 \times 10^5 {\rm M}_{\odot}$
for the total number of stars ($N_{\rm s}$)  being $10^6$.
We also investigate the models with higher $m_{\rm l}$ in order to
investigate GCs with higher $M_{\rm gc}$ for a given $N_{\rm s}$.
We particularly investigates the models with $m_{\rm l}=0.3 {\rm M}_{\odot}$ and
$N_{\rm s}=10^6$ (corresponding to $M_{\rm gc}=8.1 \times 10^5 {\rm M}_{\odot}$)
We particularly investigates the models with $m_{\rm l}=0.8 {\rm M}_{\odot}$ and
$N_{\rm s}=1.3 \times 10^6$ 
(corresponding to $M_{\rm gc}=2.3 \times 10^5 {\rm M}_{\odot}$).

\subsection{Gaseous  distributions}

Bekki (2010, B10) demonstrated that gas ejected from AGB stars of 1G populations
can finally form a  gas disks in the central region of the original
stellar system, if 1G stars initially have a small amount of rotation: this
is confirmed by B11.  
Our previous and recent simulations showed
that gas accreted from a giant molecular cloud (GMC) 
onto a GC can form a  compact gas
disk within the GC using their new models of GMC-GC
interaction  (BM09;
McKenzie and Bekki 2018) 
It is thus reasonable for the present study to assume a gas disk embedded
in a DSS. 
In the present paper, we investigate exclusively the models in which
the initial gas disks have uniform distributions with different masses ($M_{\rm g}$) and sized ($R_{\rm g}$).

A gas disk is represented by equal-mass gas
particles with 
and each particle is assumed to isothermal with its temperature
${\rm T}_{\rm g}$ of 10 K
for most models: we 
do not include the time evolution of 
$T_{\rm g}$ of  these initially cold gas particles,
because any feedback effects associated with star formation (e.g.,
massive stars and supernovae) are not considered in the present study.
This could be over simplified in gas dynamics within GCs, because
AGB ejecta with temperature of more than 1000 K could heat up the existing
gas. D08 showed that AGB ejecta can be cooled down to be accumulated into
the central regions of GCs, though their simulations are based on 1D
models (i.e., spherically symmetric distributions of gas and stars).
It is beyond the scope of this paper to investigate how new AGB ejecta
can influence the existing gas disk during GC evolution.
For most models, the mass of each gaseous particle is set to be 
$0.01 {\rm M}_{\odot}$. 
Accordingly, a gas disk with $M_{\rm g}=10^4 {\rm M}_{\odot}$ consists
of $10^6$ SPH gas particles in the present study.

The gas disk in a GC is assumed to be rotating within GCs, which is consistent
with our previous simulations (B10, B11). Accordingly, $i$th gas particle
has a circular velocity ($v_{\rm c, \it i}$) determined by
the gravitational potential of the GC and the gas disk as follows:
\begin{equation}
v_{\rm c, \it i}=g(\phi (x_i)),
\end{equation}
where $x_i$ is the 3D position of the particle with respect to the GC's center,
$g$ is a  function that determines $v_{\rm c, \it i}$ from the gravitational
potential $\phi$.
Recent observations have revealed that the Galactic GCs have a high degree of
rotation with $V/\sigma$ ranging from 0.1 to 0.5 (e.g., Bianchini et al. 2018).
This  means that the initial degree of rotation in GCs should be significantly higher
than the above, because two-body relaxation effects reduce the degree
(Bianchini et al. 2018). Milone et al. (2018a) have demonstrated that
stellar kinematics of 2G stars in 47 Tuc shows a higher degree of anisotropies and 
smaller tangential velocity dispersion than 1G stars and suggested that
these are consistent with a formation scenario in which 2G stars originate
from a compact configuration with strong rotation
(e.g., Mastrobuono-Battisti \& Perets 2013). These recent observations justify
the adopted compact gas disks with rotation in the present study.

\begin{figure}
\psfig{file=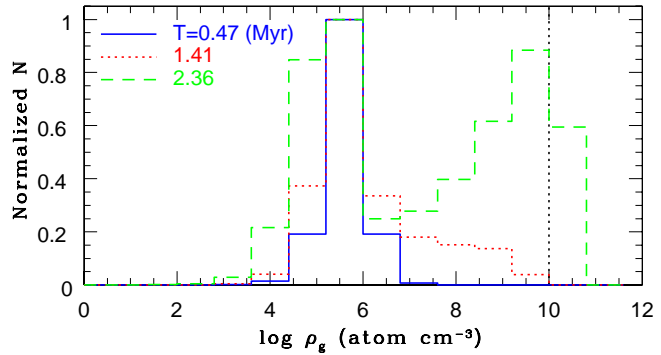,width=8.5cm}
\caption{
Distribution of gas densities ($\rho_{\rm g}$)
for the fiducial model MN2 without star formation at three  different time steps,
 $T=0.81$ Myr, 1.24 Myr, and 2.35 Myr.
Normalized number of gas particles is given for each density bin.
}
\label{Figure. 4}
\end{figure}

\begin{figure*}
\psfig{file=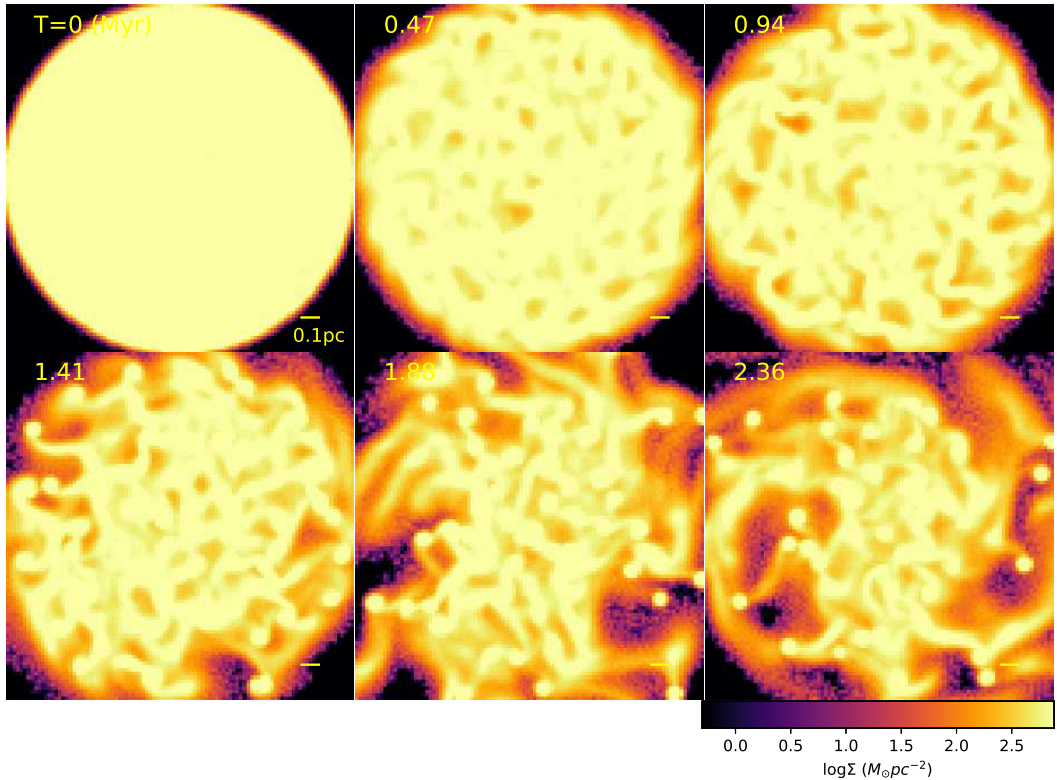,width=18.0cm}
\caption{
The same as Fig. 1 but for the model MN1 with lower gas mass fraction
($f_{\rm g}=0.003$)
}
\label{Figure. 5}
\end{figure*}

\subsection{Star formation}

Our previous simulations have already investigated
star formation from gas in fractal GMCs that can form
GCs with both 1G and 2G stars (B17b).
The spatial resolution of the simulation is high enough
to identify high-density gas with the mass density 
of $\sim 10^5$ atom cm$^{-3}$,
which corresponds to cores of GMCs. In the present study,
we consider that if the gas density of a particle can become
higher than the typical density of first stellar cores 
($10^{10}$ atom cm$^{-3}$),
then a new star can be formed. 
Therefore, the present simulations can resolve the formation of 
each individual star,  which is a main different between the present study
and B17b.

When the mass density of a gas particle
exceeds a threshold value for
star formation ($\rho_{\rm g, th}$) in  a model with star formation,
then the particle is
converted into a  collisionless
star particle. Accordingly, the present star formation model is as follows:
\begin{equation}
\rho_{\rm g} > \rho_{\rm g, th},
\end{equation}
where $\rho_{\rm g,th}$ is set to be $10^{10}$ atom cm$^{-3}$, which corresponds
to the mass  density of first stellar cores (e.g., Meyer 1978).
The mass of the new stellar particle ($m_{\rm ns}$) is always the same
as $m_{\rm g}$ of its original gas particle, because no gas ejection
from the new star is included in the present study.
We investigate the star formation of a model over 2.4 Myr in the present
study. We however confirm that star formation becomes very small after
$\sim 3$ Myr owing to rapid consumption of gas by star formation
in a model with $\sim 30$ Myr evolution.
This means that the long-term evolution of gas and star
formation is not so important in discussing the star formation process
within GCs in the present models.

\subsection{Representative models}

We investigate star formation processes within GCs for models with
different $M_{\rm gc}$,
$R_{\rm gc}$ ($=5a_{\rm gc}$),
$M_{\rm g}$, and $R_{\rm g}$.
In order to demonstrate the effects of SB on secondary star formation
more clearly, 
we also investigate the models with and without SB. In the models without
SB, the gravitational potentials of the GCs are fixed (i.e., no stellar motion)
during the simulations. 
We also investigate the models in which star formation is not included,
in order to demonstrate the effects of SB on the formation of high-density
gas clumps more clearly. These models without SF are labeled as ``MN'' 
(e.g., MN1)
whereas those with SF are labeled as ``M'' (e.g., M1) 

The model MN2 with $M_{\rm gc}=3.1 \times 10^5 {\rm M}_{\odot}$,
$R_{\rm gc}=2$ pc, $M_{\rm g}=3000 {\rm M}_{\odot}$ (and without 
star formation) is denoted as the fiducial model.
This model and other MN models are investigated in detail,
because the effects of SB on the mass growth of small 
(star-forming) gas clumps are crucial in the present study.
The basic model parameters for the fiducial model are given in Table 2.
We mainly investigate the models with star formation and
$M_{\rm gc}=3.1 \times 10^5 {\rm M}_{\odot}$, which is similar to
the present-day typical GC mass.
It should be noted that $M_{\rm gc}$ is the total mass of stars and
compact objects (stellar mass black holes and neutron stars) in a GC
after explosions of all SNII. 

We also investigate star formation
processes in  high-mass GC models in which
$M_{\rm gc}$ is as large as or larger than $10^6 {\rm M}_{\odot}$,
because previous theoretical models predicted that original GC masses
should be at least $[5-10]$ times larger than the present-day typical
GC mass (e.g., Bekki \& Norris 2006; D08).
In these models, the lower-mass cut-off of the IMF ($m_{\rm l}$)
is set to be larger than $0.1 {\rm M}_{\odot}$ so that 
$M_{\rm gc}$ can be larger for a given IMF slope and a total number
of stellar particles.
The values of physical parameters for all models are summarized in Table 3.

\subsection{Resolution issues}

Previous studies investigated resolution requirements for accurate simulations of the formation of collapsed objects in circumstellar disks
(e.g., Bate \& Burkert 1997; Truelove et al. 1997; Nelson 2007). 
Although these study are not directly related to the present study,
their results are useful in discussing whether the new results in
the present study are due to numerical artifacts. 
The Jeans mass ($M_{\rm J}$) in the model M4
with $M_{\rm g}=3000 {\rm M}_{\odot}$ is $3.7 {\rm M}_{\odot}$ 
at $T=0$, and it is much larger than the mass resolution 
of the simulation ($0.01 {\rm M}_{\odot}$). $M_{\rm J}$ is even higher
for models with lower $M_{\rm g}$: it is $11.7 {\rm M}_{\odot}$ for
$M_{\rm g}=300 {\rm M}_{\odot}$. Therefore, the present simulations
can properly investigate the formation of gas clumps in a gas disk
embedded in a dense stellar system.

Truelove et al. (1997) defined the ``Jeans number'' ($J$) as follows:
\begin{equation}
J = \frac{ \Delta x }{ {\lambda}_{\rm J}  }  \;
\end{equation}
where $\Delta x$ and ${\lambda}_{\rm J}$ are the cell size of a simulation
and Jeans length, respectively. They thereby demonstrated that
if $J$ is kept lower than 0.25, then artificial fragmentation of a gas cloud
can be avoided. 
The present simulation code is quite different from those adopted
in Truelove et al. (1997). 
We therefore redefine $J$ as follows:
\begin{equation}
J = \frac{ \epsilon_{\rm g} }{ {\lambda}_{\rm J}  }  \;
\end{equation}
where $\epsilon_{\rm g}$ is the gravitational softening length of gas
particles (corresponding to spatial resolution). 
The Jeans number $J$ is estimated to be 0.044 for models with
$M_{\rm g}=3000 {\rm M}_{\odot}$ at $T=0$ and it is lower
in the models with lower $M_{\rm g}$  in the present study.
Therefore, the formation of high-density clumps is not due to
artificial fragmentation caused by the discreteness in the initial
gas distributions of simulations.

\begin{figure}
\psfig{file=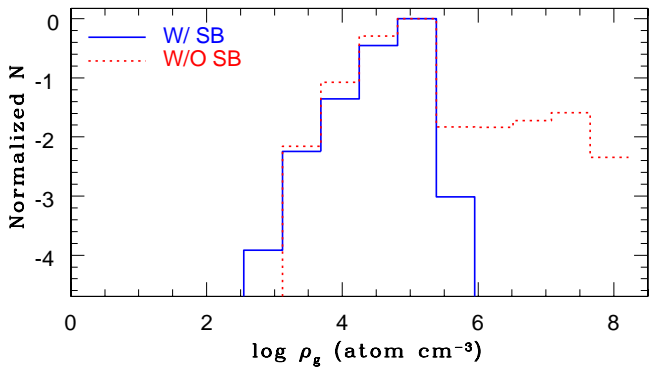,width=8.5cm}
\caption{
The same as Fig. 4 but for the model MN1 with SB and MN4 without SB
at $T=0.94$ Myr.
}
\label{Figure. 6}
\end{figure}

\begin{figure*}
\psfig{file=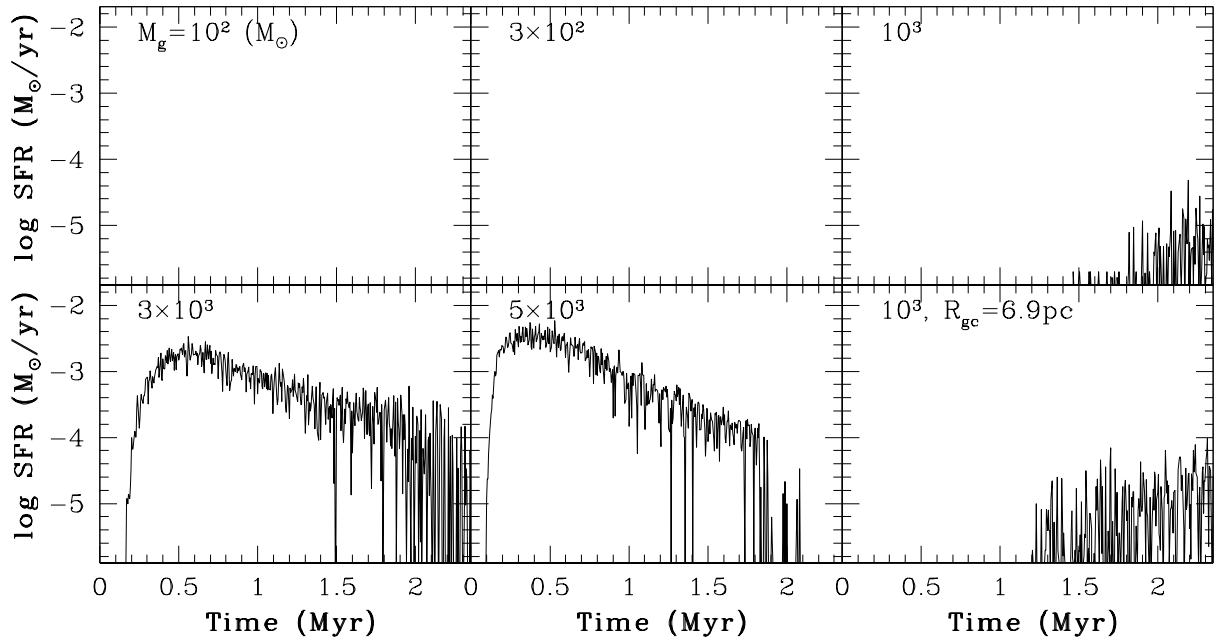,width=16.0cm}
\caption{
Time evolution of star formation rates for six different models with
$M_{\rm gc}=3.1 \times 10^5 {\rm M}_{\odot}$. The initial GC size ($R_{\rm gc}$)
is the same between the models 
($R_{\rm gc}=10$pc) except the model shown in the lower right (6.9pc).
The initial gas mass is shown in the upper left corner of each panel.
}
\label{Figure. 7}
\end{figure*}

\begin{figure*}
\psfig{file=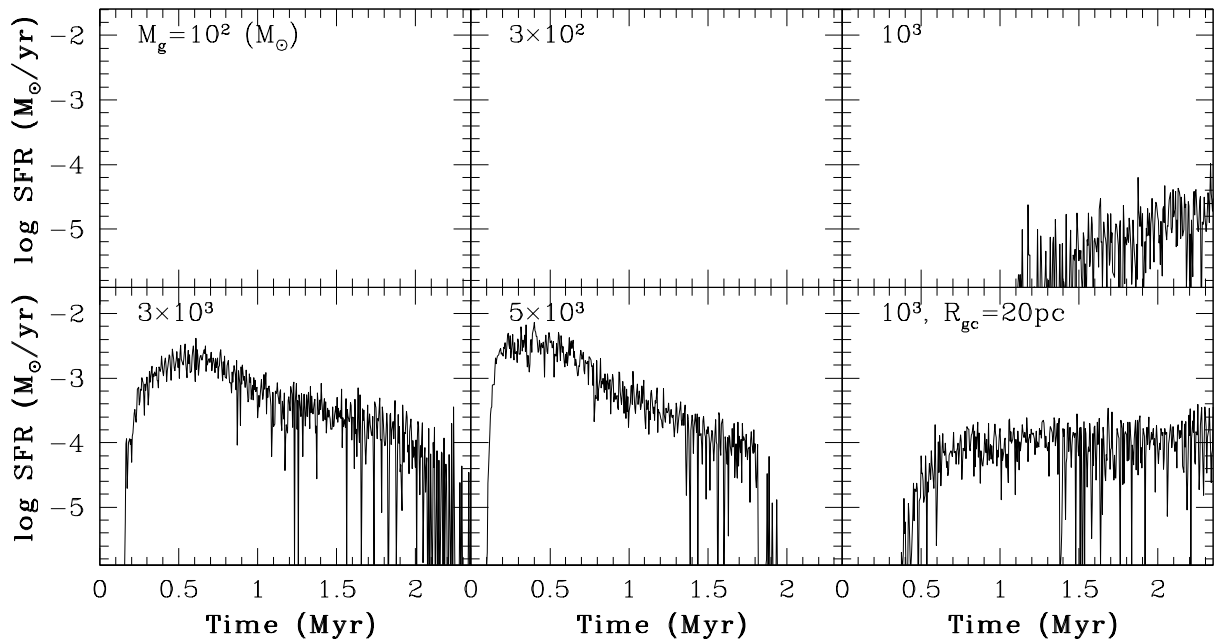,width=16.0cm}
\caption{
The same as Fig. 6 but for $M_{\rm gc}=1.5 \times 10^5 {\rm M}_{\odot}$.
}
\label{Figure. 8}
\end{figure*}

\begin{figure*}
\psfig{file=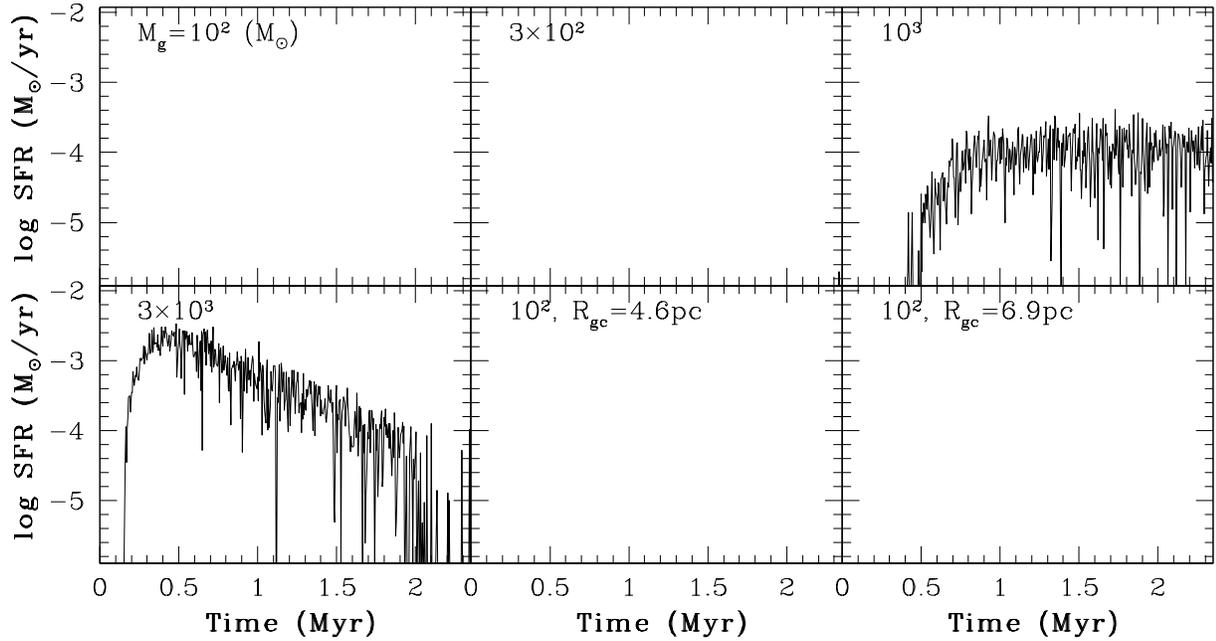,width=16.0cm}
\caption{
The same as Fig. 6 but for $M_{\rm gc}=3.1 \times 10^4 {\rm M}_{\odot}$.
Two models with different $R_{\rm gc}$ are shown.
}
\label{Figure. 9}
\end{figure*}

\begin{figure*}
\psfig{file=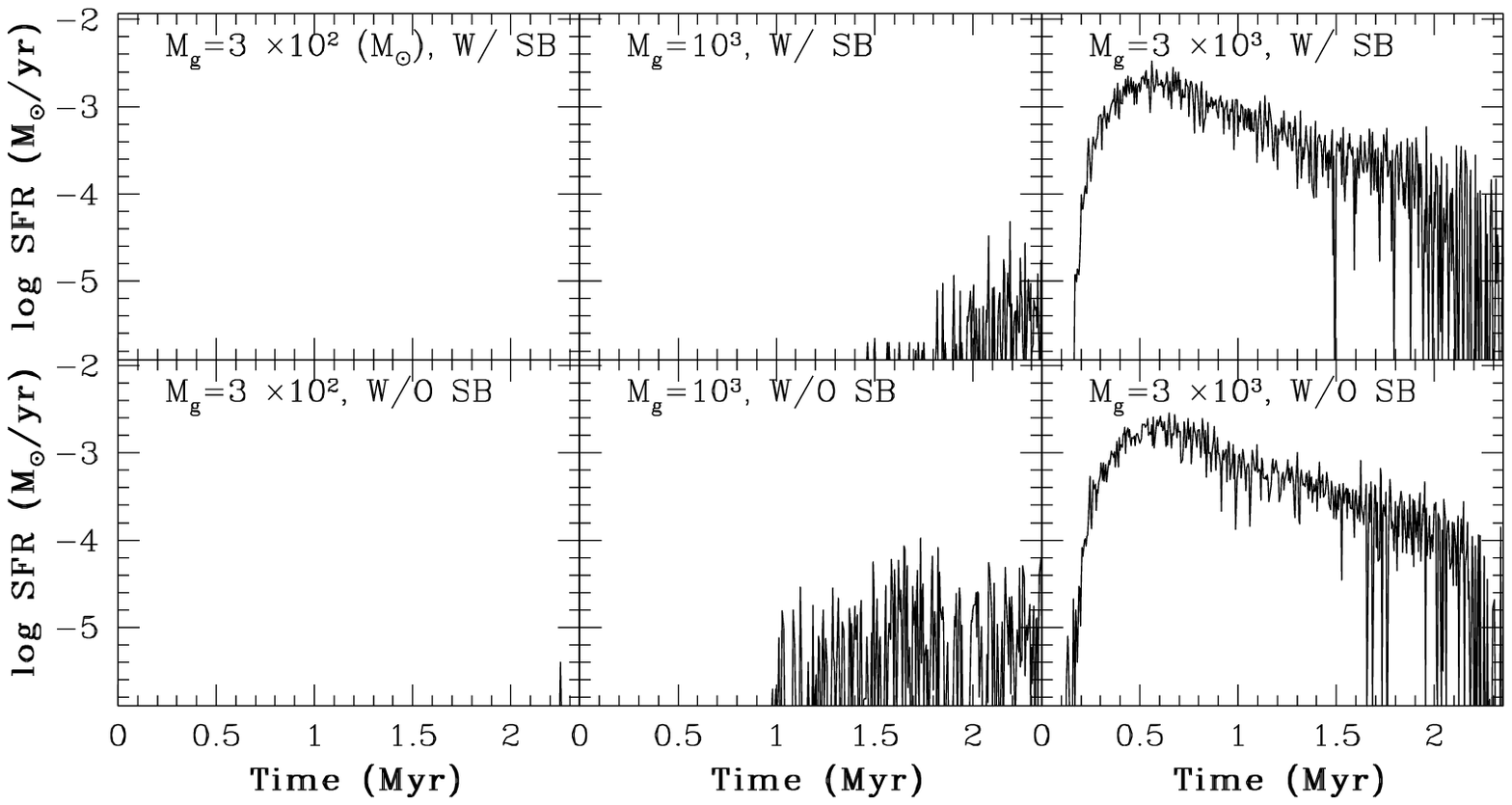,width=16.0cm}
\caption{
The same as Fig. 6 but for the models with SB (upper three)
and without SB (lower three). In these models,
$M_{\rm gc}=3.1 \times 10^5 {\rm M}_{\odot}$ and $R_{\rm gc}=10$pc.
}
\label{Figure. 10}
\end{figure*}

\section{Results}

\subsection{Effects of stellar bombardment (SB)}

Fig. 2 shows how high-density gas clumps can be developed during the 
evolution of a gas disk embedded in a GC for a fiducial model
MN2 with $M_{\rm gc}=3.1 \times 10^5 {\rm M}_{\odot}$ and $a_{\rm gc}=2 $pc
($R_{\rm gc}=10$pc), and $M_{\rm g}=3 \times 10^3 {\rm M}_{\odot}$, and 
$R_{\rm g}=1$ pc.
Clearly, numerous small gas clumps with
the surface densities as high as $10^{5}$ ${\rm M}_{\odot}$ pc$^{-2}$ 
can be developed from local gravitational instability around $T=1.4$ Myr.
The sizes of these high-density clumps appear to be similar in this figure,
and they do not merge frequently with one another to form more massive clumps.
Dynamical friction of these clumps against GC member stars cannot be efficient
within a timescale of $10^6$ yr, which ends up with no strong concentration
of gas in the GC's center. As shown in Fig. 3, a few gas clumps cannot stay in the 
original thin gas disk, because they interact with other gas clumps and with
individual stars intruding into the gas disk from various directions.

Using the stability criteria for a uniformly rotating isothermal disk
(Goldreich \& Lynden-Bell 1965),
the origin of the dense gas clump formation described above 
can be discussed as follows.
The stability parameter for gas ($Q_{\rm g}$) id described as follow:
\begin{equation}
Q_{\rm g}= \frac {v_{\rm s} \Omega}{ G \Sigma_{\rm g} } \;
\end{equation}
where $v_{\rm s}$, $\Omega$, $G$, and $\Sigma_{\rm g}$ are 
the sound velocity ,
angular speed, gravitational constant, and surface
density of gas, respectively.
If this $Q_{\rm g}$ is less than 1.06 in a gas disk, then
the disk becomes unstable.
It is found that (i) $Q_{\rm g}$ ranges from 1.04 ($r=1$ pc)  to 1.64
(0.1 pc) in this model
and (ii) it is lower than $1.06$ in the outer part of the 
disk.
This means that the gas disk is marginally stable against local
disk instability.
However, the formation of dense gas clumps
(and the subsequent star formation)  is possible due to local
instability in the regions with lower $Q_{\rm g}$:
clump formation 
can  proceed more efficiently in the outer region with lower $Q_{\rm g}$.
The models with lower $M_{\rm g}$ ($\le 1000 {\rm M}_{\odot}$) have
higher $Q_{\rm g}$ ($>2$) so that dense gas clumps cannot be efficiently
formed from local instability within the gas disks.

Fig. 4 describes how high-density gas clumps can be gradually developed
in the gas within a timescale of Myr. Only a small fraction of gas particles
can have gas densities ($\rho_{\rm g}$)
larger than $10^{10}$ atom cm$^{-3}$ corresponding
to first stellar cores (i.e., a threshold gas density for star formation)
at $T=1.41$ Myr. A significant fraction of gas particles
can finally  have $\rho_{\rm} \sim 10^{10}$ atom cm$^{-3}$ at $T=2.36$ Myr,
which implies that star formation cannot start soon after gas accretion
from AGB stars or ISM onto GCs. It should be noted here that most of previous
models for multiple stellar populations of GCs (e.g., B07, D08, BJP17)
assumed star formation soon after gas accretion in GCs (i.e., no time delay
between gas accretion and star formation is assumed).

It should be noted here that no initial turbulence is introduced in the 
original gas disks in these models. Therefore, the formation of 
high-density gas clumps and the subsequent star formation in these
models should be quite different from those in fractal gas clouds
with initial turbulence investigated in previous studies
(e.g., Klessen et al. 1998; Dale 2011). 
Therefore,  the present study, for the first time,
has demonstrated that star formation is possible from gas without
turbulence. 
As described in later for the models with star formation,
a certain fraction of gas particles can have 
$\rho_{\rm g} \ge 10^{10}$ atom cm$^{-3}$ at 0.47 Myr,
which means that hydrodynamical pressure of gas can suppress
the mass growth of gas clumps in the MN models without SF.

Fig. 5 shows the formation of gas clumps in the model MN1 in which
all model parameters except $M_{\rm g}$ are the same as those adopted
in  MN2.  Clearly, the surface
densities ($\Sigma_{\rm g}$) of gas clumps in this model
with lower $M_{\rm g}$ are much (by almost two orders of magnitude) lower
than those in MN2, which implies that $f_{\rm g}$ is a key factor which
determines whether star formation is possible.
Star formation can be significantly delayed or not be possible in
GCs with lower $f_{\rm g}$ ($<0.01$).
This possible threshold $f_{\rm g}$ for star formation can have several
important implications on the origins of multiple stellar populations,
as discussed later in  this paper.

In order to demonstrate  the negative effects of SB on secondary star formation
more clearly, we have investigated models without SB.
Fig. 6 describes how SB can influence the formation of high-density clumps
in the early evolution of gas in a GC for the models MN1 with SB
and MN4 without SB.  The model with SB does not show any gas particles
with $\rho_{\rm g}>10^6$ atoms cm$^{-3}$, which means that the mass growth
of gas clumps can be severely suppressed in this model with lower 
$M_{\rm g}$ ($=1000 {\rm M}_{\odot}$). However, there are no significant
differences in the normalized distributions of $\rho_{\rm g}$ between
the two models for $\rho_{\rm g}<10^5$ atom cm$^{-3}$. 
These results clearly demonstrate that star formation processes such as
the mass fraction of gas converted into new stars and the timescale
of star formation can be influenced by SB when $f_{\rm g}$ is low.
It is confirmed that this SB effect is not seen in the models with
$f_{\rm g} \ge 0.01$.

\begin{table}
\centering
\begin{minipage}{85mm}
\caption{Star formation efficiencies ($\epsilon_{\rm sf}$) 
for  the representative models with different
$f_{\rm g}$ and $R_{\rm g}$.
}
\begin{tabular}{llll}
$M_{\rm gc}$ (${\rm M}_{\odot}$)
& $R_{\rm g}$ (pc) & $f_{\rm g}$ & $\epsilon_{\rm sf}$ \\
$3.1 \times 10^5$ & 1  & 0.01  & 0.73 \\
$3.1 \times 10^5$ & 2  & 0.01  & 0.30 \\
$3.1 \times 10^5$ & 3  & 0.01  & 0.12 \\
$3.1 \times 10^5$ & 3  & 0.1  & 0.47 \\
$8.1 \times 10^5$ & 1  & 0.006  & 0.70 \\
$8.1 \times 10^5$ & 3  & 0.03  & 0.36 \\
$8.1 \times 10^5$ & 3  & 0.1  & 0.49 \\
$2.3 \times 10^6$ & 1  & 0.002  & 0.46 \\
$2.3 \times 10^6$ & 3  & 0.005  & 0.07 \\
$2.3 \times 10^6$ & 3  & 0.05  & 0.45 \\
\end{tabular}
\end{minipage}
\end{table}

\begin{figure*}
\psfig{file=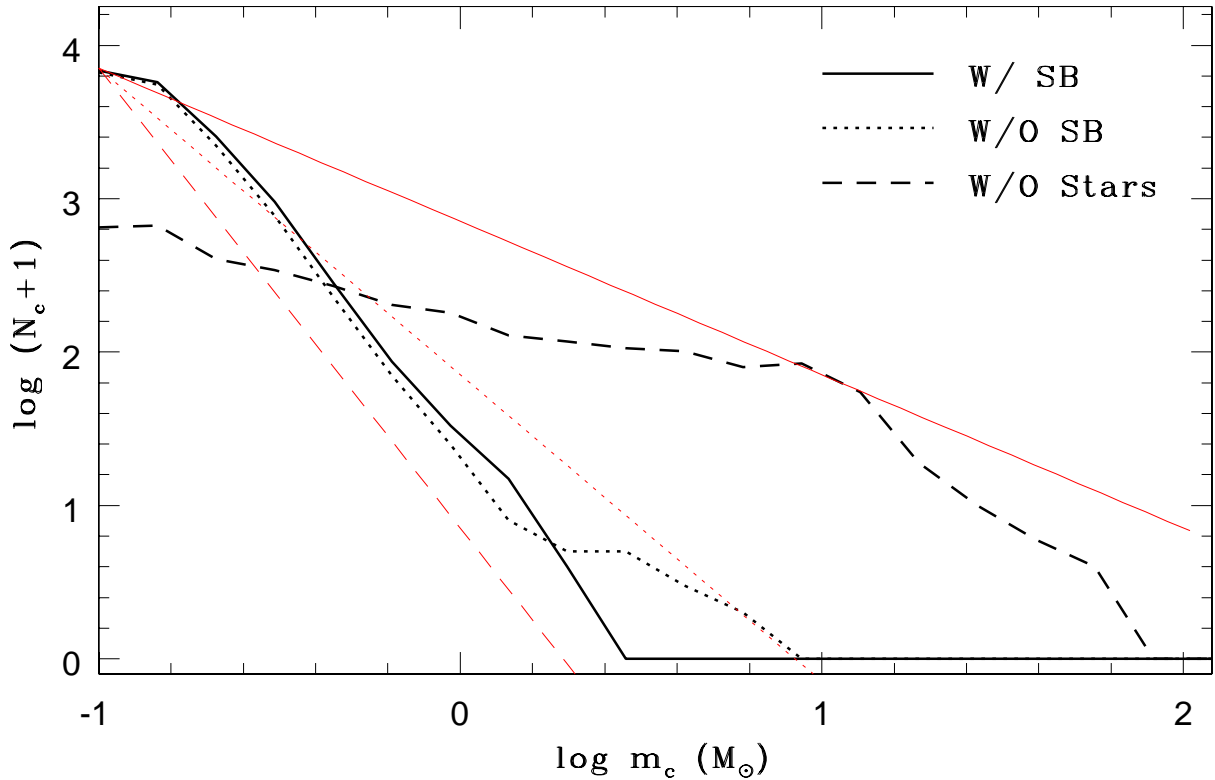,width=16.0cm}
\caption{
Number distributions ($N_{\rm c}$)
of gas cloud masses ($m_{\rm c}$) for the models
with $M_{\rm gc}=3.1 \times 10^5 {\rm M}_{\odot}$ 
and $f_{\rm g}=0.01$ (black solid line)
at $T=0.47$ Myr.
For comparison,
the same models yet without SB (dotted) and without stars and compact
stellar objects (dashed) are shown.
The model labeled as ``W/O stars'' accordingly means that there is no global
gravitational field of a GC.  For convenience, $\log (N_{\rm c}+1)$
is shown in this figure. The three slopes of the cloud mass functions,
$\alpha_{\rm c}=-1$ (red solid), $-2$ (dotted), and $-3$ (dashed) are
also shown for comparison.
}
\label{Figure. 10}
\end{figure*}

\begin{figure*}
\psfig{file=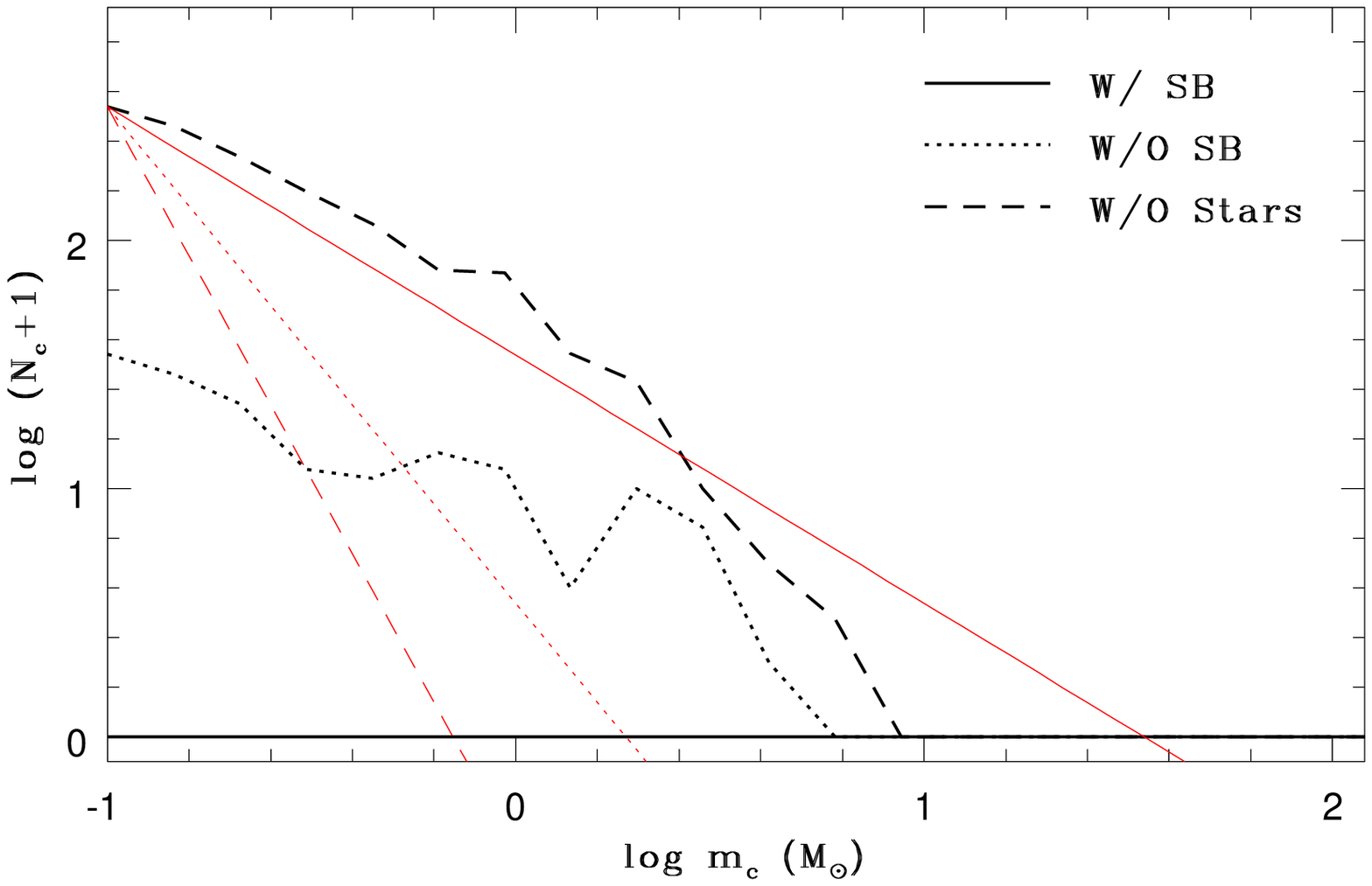,width=16.0cm}
\caption{
The same as Fig. 11 but for the models with $f_{\rm g}=0.003$ at
$T=0.94$ Myr. No gas clouds that are collapsing and have masses
more than $0.1 {\rm M}_{\odot}$ are found in the model with SB
(i.e., solid line).
}
\label{Figure. 11}
\end{figure*}

\subsection{Star formation in different models}

Fig. 7 describes how the star formation histories within GCs
depends on $M_{\rm gc}$ and $R_{\rm gc}$ in 
the models M1-M6.
Secondary star formation is either completely shut down
or very severely suppressed due to SB and gravitational
potentials of existing stars in the models with $f_{\rm g} <0.01$
(i.e., M1, M2, and M3), and this effect can be also seen in the model
M6 with a very compact GC. These results confirm that $f_{\rm g}$ 
is the most important parameter for secondary star formation.
The model M4 with $f_{\rm g}=0.01$ shows a bursty star formation
with $\sim 80$\% of the initial gas being converted into new stars
(i.e., $\epsilon_{\rm sf}=0.8$)  within
$\sim 2$ Myr. This timescale of star formation 
is shorter than the lifetime ($\sim 3$ Myr)
of the most massive stars adopted in
the present study ($120 {\rm M}_{\odot}$).
Such short and bursty star formation from gas within GCs
ensures that secondary star formation cannot be interrupted
by SNII before the consumption of most of the gas.
These results demonstrate that there is a threshold gas fraction
($f_{\rm g, th}$) above which short and  bursty
star formation is possible.

Fig. 8 confirms that GCs with lower $f_{\rm g}$ do not show
short bursty star formation in the models with 
$M_{\rm gc}=1.5 \times 10^5 {\rm M}_{\odot}$. However,
the model M9 with lower GC masses shows weak star formation
for $f_{\rm g}=0.006$, which is a hint for a higher $f_{\rm g, th}$.
The model M12 with lower $f_{\rm g}$ and larger $R_{\rm gc}$ 
(thus less compact) shows significantly more star formation
than M9 with the same $f_{\rm g}$ yet smaller $R_{\rm gc}$.
This more efficient star formation in less compact stellar systems
implies that the degree of self-gravitation in gas (i.e.,
$M_{\rm g}/M_{\rm gc}$)  within $R_{\rm gc}$ is crucial for
star formation in GCs. For this model M12,  star formation
can continue with an almost constant star formation
rate for a longer timescale due to the lower
star formation efficiency.

Although Fig. 9 confirms that there is $f_{\rm g, th}$ in GCs with 
and $M_{\rm gc}=3.1 \times 10^4 {\rm M}_{\odot}$, 
the model with $f_{\rm g}=0.01$ ($M_{\rm g}=3 \times 10^2 {\rm M}_{\odot}$)
does not show any star formation, which is in a striking contrast with M4 
with $f_{\rm g}=0.01$ that shows short bursty star formation.
Although star formation is relatively active in the model with $f_{\rm g}=0.03$,
it proceed gradually with an almost constant rate without burst.
This is quite different from the more massive  models with $f_{\rm g}=0.03$
and $M_{\rm gc}=3.1 \times 10^5 {\rm M}_{\odot}$.
The model with $f_{\rm g}=0.1$ shows a similar short bursty behavior like M8,
which demonstrates that $f_{\rm g, th}$ should be larger for GCs with 
lower masses.
It should be noted here, however, that $f_{\rm g}=0.1$ is possible
only if almost all AGB ejecta is accumulated in GCs for a canonical
(Salpeter) IMF (see Fig. 1 in B11).
Therefore, it is reasonable to say that low-mass GCs are unlikely
to show short bursty star formation from gas accumulated from
AGB stars within the GCs.
If such low-mass GCs show short bursty star formation, then gas should 
be supplied from outside GCs (e.g., ISM).

GCs with $f_{\rm g}=0.003$ do not show star formation in the models
with $M_{\rm gc}=3.1 \times 10^4 {\rm M}_{\odot}$ irrespective of
$R_{\rm gc}$, as shown in Fig. 9.
This suggests that the gas density above which star formation can star
within GCs is higher for low-mass GCs.
These results shown in Figs. 7-9 suggest that secondary star formation
can start earlier (i.e., when $f_{\rm g}$  is lower)
in more massive GCs. These also suggest that efficient star formation
is not possible in low-mass GCs with $M_{\rm gc}<10^5 {\rm M}_{\odot}$,
even if gas can be accumulated in the central regions of the GCs.
It should be noted here that these low-mass GCs are less likely
to retain AGB ejecta owing to the lower escape velocities (B11).

Fig. 10 demonstrates that GCs show lower star formation rates
in the model M3 with SB than in M32 without SB for
$f_{\rm g}=0.003$. This suppression of star formation due to SB
can be barely seen in the models M2 with SB
and M31 without SB for $f_{\rm g}=0.001$.
For these models with low $f_{\rm g}$,
global potentials of GCs alone  can suppress star formation.
Very low star formation in the model
without SB for low $f_{\rm g}$ means that the threshold $f_{\rm g}$ for
the onset of star formation is lower for the models without SB. 
The models M4 and M33 with and without SB, respectively, do not show
any significant differences in their star formation histories,
which demonstrates that SB is not important for GCs with higher $f_{\rm g}$.
In these models,  the larger degree of self-gravitation in gas can cause
more rapid growth of high-density gas clumps within GCs so that star-gas
direct interaction cannot greatly influence the formation and evolution
of gas clumps.
Thus, secondary star formation within GCs can be influenced by SB only when
$f_{\rm g}$ is low ($<0.01$): SB can influence the star formation process
only in the early phase of GCs with gas.

Table 4 summarizes $\epsilon_{\rm sf}$ for more massive GC models with
$M_{\rm gc}$ ranging from $3.1 \times 10^5 {\rm M}_{\odot}$
to $2.3  \times 10^6 {\rm M}_{\odot}$. First, it is clear that
$\epsilon_{\rm sf}$ is quite high ($>0.4$) for all models
with $f_{\rm g} \ge0.01$ and $R_{\rm g}=1$ pc, which suggests that star formation can proceed
in a bursty manner even in the earlier phases of gas accretion within GCs.
Second, massive GCs with $M_{\rm gc} \ge 8.2 \times 10^5 {\rm M}_{\odot}$ 
show $\epsilon_{\rm sf} >0.7$ even for $f_{\rm g}=0.006$. Since,
the mass ratio of gas from massive AGB stars
with $m_{\rm s} \ge 7 {\rm M}_{\odot}$ to the original GC mass
is about 0.008 for a canonical IMF
(adopted in this study) and a mass fraction of AGB ejecta (lost to ICM)
being 0.8,
this result means that if
ICM originates exclusively from AGB stars, then star formation is possible
directly from gas ejected from massive super AGB stars with masses 
larger than $7 {\rm M}_{\odot}$.

Third, $\epsilon$ can be lower in the models
with larger $R_{\rm g}$ (2pc and 3 pc) which correspond to GCs with
a larger amount of angular momentum in their gaseous components.
Although $\epsilon_{\rm sf}$ is not so low ($\sim 0.3$),
$f_{\rm g}$ should be significantly larger for larger $R_{\rm g}$ for high 
$\epsilon_{\rm sf}$ ($>0.5$).
For example, $\epsilon_{\rm sf}$ is $\sim 0.5$ even if $f_{\rm g}$ is
0.11 in the models 
with $M_{\rm gc}=3.1 \times 10^5 {\rm M}_{\odot}$ and $R_{\rm g}=3$ pc.
This required high $f_{\rm g}$ can be achieved only if almost all gas from
massive and intermediate-mass AGB stars can be accreted onto GCs.
This accordingly suggests that it is unlikely for larger gas disks to form
new stars with high 
$\epsilon_{\rm sf}$ ($>0.5$).
Since the gas disk sizes within GCs can be determined by the initial angular
momentum of their 1G stars (B11), 
the above result implies that angular momentum of GCs can be a key determinant
for star formation within GCs.

%

\subsection{Mass functions of gas clumps}

The models without star formation
are the best to demonstrate the effects of SB on the mass growth of small
clumps and the mass function (MF) of the clouds within GCs more clearly,
We here describe the results for two representative models with
$M_{\rm gc}=3.1 \times 10^5 {\rm M}_{\odot}$ 
and  $f_{\rm g}=0.003$
and 0.01.
As shown in Fig. 11,  the MF of gas clumps
developed from local instability within GCs is not so much different between
the models with and without SB for lower masses 
($m_{\rm c} < 3 {\rm M}_{\odot}$) at $T=0.47$ Myr,
when star formation can be very active if star formation
is included. 
However, there are no massive clumps with
$m_{\rm c} > 3 {\rm M}_{\odot}$ in the model with SB,
which suggests that the formation of more massive stars
with $m_{\rm c} \ge 3 {\rm M}_{\odot}$
can be suppressed
by SB. The slope of the MF ($\alpha_{\rm c}$) is approximately
$-2.5$ for 
$m_{\rm c} < 3 {\rm M}_{\odot}$ at $T=0.47$ Myr for the two models.

The comparative model without stars (i.e., no gravitational potential of
a GC) shows a quite flat  MF of clouds for
$m_{\rm c} < 10 {\rm M}_{\odot}$ and a significant fraction
of massive clouds with
$m_{\rm c} \ge 8 {\rm M}_{\odot}$.
The initial rotating gas disk is initially unstable ($Q_{\rm g}<1$) 
so that massive 
numerous gas clumps can be developed rapidly from local instability.
This model is introduced such that the physical roles of background
potential of stars in the formation of gas clumps (thus gas
mass function) can be more clearly shown.
These results suggest that massive star formation that leads to 
SNII can be severely suppressed by a combination
of star-gas interaction and gravitational potentials of GCs.
These furthermore imply that the IMF can be ``top-light'' in secondary
star formation within GCs and thus that SNII feedback effects on
gas left after the formation of preceding generations of stars
are less dramatic owing to the smaller number of
massive stars. If the formation of gas clumps with 
$m_{\rm c} \ge 8 {\rm M}_{\odot}$ is completely suppressed, then
secondary star formation can continue until all of the gas is consumed.

Fig. 12 shows that no high-density gas clumps with 
$m_{\rm c} \ge 0.1 {\rm M}_{\odot}$ can be formed 
at $T=0.94$ Gyr in the model with $f_{\rm g}=0.003$ and SB.
Its comparative model without SB, however, shows the formation
of small clouds with $m_{\rm c}$ up to  $\sim 6 {\rm M}_{\odot}$
and a flat mass function. Fig. 11 also confirms that 
the GC's gravitational potential can strongly suppress the formation
of massive small clumps with 
$m_{\rm c} \ge 8 {\rm M}_{\odot}$.
Furthermore, it is clear that the GC's potential 
can suppress the formation of low-mass clumps too in the model
with this low $f_{\rm g}$, in which the degree of gaseous self-gravity
is so low that gas clouds cannot grow rapidly through accretion
of nearby gas.
Thus, the results for the models with $f_{\rm g}=0.003$ and 0.01 demonstrate
that the formation of massive stars that can explode as SNII can be
severely suppressed, in particular, in the early phases of GCs
with lower $f_{\rm g}$.

\section{Discussion}

\subsection{Limitation of the models}

The present study has assumed that gas accreted from AGB stars
and/or ISM can have a disky structure and rotation within a GC.
Although such a rotating gas disk has been demonstrated to be formed
in recent simulations of GC formation (e.g., B11; BM09), it is still
possible that the adopted gas disk model is over-simplified and less
realistic as follows.
After the formation of a disk from gas ejected from AGB stars in a GC
(i.e., the formation of ``existing disk''),
a significant amount of 
ISM can be accreted onto the GC. The spin
of the existing gas disk consisting of AGB ejecta would not be necessarily
alligned with that of  ISM accreting onto the GC, because gas accretion
can be possible from any direction. The alignment of spin axes between
the existing gas disk and the accreting ISM
is possible, only if the spin axis of the GC is well aligned with 
the gas disk of the GC-hosting galaxy.

If the angular momentum vector of the accreting ISM is quite different from
that of the existing  gas disk, then such ISM accretion could 
significantly change the structure and kinematics of the existing gas disk.
Accordingly, the gas disk that is assumed to be thin and rotating
over $\sim 3$ Myr in the present study
could be over-simplified.
If dynamical and hydrodynamical interaction between the existing gas disk
and accreting ISM can heat up the disk (i.e., high gaseous temperature),
then $Q_{\rm g}$ becomes significantly larger ($>1$), which ends up with
suppression of star formation. Also, such interaction transforms the
initially  thin
disk to a considerably thick one, and then star formation could be suppressed too
owing to the lower surface gas density.
It is our future study to investigate how the structure and kinematics of
gas originating from AGB ejecta and ISM within GCs can evolve with time for
models with different spin axes of GCs with respect to spin axes
of their host
galaxies' gas disks.
Although such a future study requires modeling both for the evolution
of galactic gas disks and for gas accretion onto GCs,
it can improve our understanding of secondary star formation within
GCs.

\subsection{Origin of He-rich stars in massive GCs}

A few massive Galactic GCs (e.g., NGC 2808 and  $\omega$Cen) are observed
to have high He abundances (Piotto et al. 2005), though some other GCs show such He
abundance enhancement to a lesser extend (e.g., Milone et al. 2017). 
A key questions related to the origin of these He-rich stars is that if they
are formed from AGB ejecta with high $Y$, 
abundance enhancement to a lesser extend (e.g., Milone et al. 2017). 
A key questions related to the origin of these He-rich stars is that if they
are formed from AGB ejecta with high $Y$, 
A key questions related to the origin of these He-rich stars is that if they
are formed from AGB ejecta with high $Y$, 
A key questions related to the origin of these He-rich stars is that if they
are formed from AGB ejecta with high $Y$, 
then external pristine gas with normal $Y$ ($\sim 0.245$)
for low-metallicity GCs) should not mix with the ejecta so that
$Y$ can be kept still high for the new stars (i.e., dilution cannot occur).
It has been, however, physically unclear why such star formation from gas
that is not mixed with pristine gas is possible
in massive GCs.
Since only massive AGB stars ($m \ge 5 {\rm M}_{\odot}$) are predicted
to eject gas with high $Y$ (e.g., Ventura \& D'Antona 2009; Karakas 2010),
star formation should start after the AGB ejecta is accumulated
yet before ejecta with lower $Y$ from low-mass AGB stars is accumulated
(to dilute the He-rich gas from massive AGB stars).

As shown in the present study,
star formation can proceed very efficiently
with $\epsilon_{\rm sf} > 0.5$ in  massive GCs, if the gas mass 
fractions ($f_{\rm g}$)
exceed 0.01. 
This $f_{\rm g, th}$ of $\sim 0.01$ is
smaller  than  the mass fraction  of gaseous ejecta
from AGB stars with $m_{\rm s} \ge 5 {\rm M}_{\odot}$ for a canonical
IMF. Therefore, star formation directly from gas from massive AGB stars
is possible.
Furthermore, such $f_{\rm g,th}$
for star formation is found to be  higher for GCs with lower masses in
the present study.
Since $f_{\rm g}$ can increase with time due to contributions of gas from
AGB stars with different masses (thus different lifetimes),
these two results imply that more massive GCs are more likely to start forming
2G stars earlier.
It is therefore possible that 2G stars with high $Y$ are formed from gas from massive
AGB stars only in massive GCs.
The present study did not investigate the gas accumulation processes of
AGB stars with different masses and the subsequent star formation in
a self-consistent manner. Thus, it is our future study to provide
more quantitative predictions on the mass fractions of He-rich stars
in GCs with different masses.


Although the origin
of the observed discrete He-rich stellar populations (2G, 3G etc)
can be understood in the context of star formation
directly from AGB ejecta,
recent observations have shown that even 1G stars can possibly
have He abundance spreads ($\delta Y$)
 to a lower degree (e.g., Millone et al. 2018b).
The average $\delta Y$ of $\sim 0.05$ is significant,
though it is smaller than $\delta Y$ observed in NGC 2808
and $\omega$ Cen. If the observed $\delta Y$ is due to
the initial $\delta Y$ of 1G stars (i.e., not due to
stellar evolution), the AGB scenario alone cannot simply
explain it. Our previous numerical simulations of GC
formation within fractal GMCs showed that stellar winds from
massive OB stars can chemically pollute GC-hosting GMCs, which ends
up with abundance spreads in He, C, N, and O among 1G stars formed
within the GMCs (Bekki \& Chiba 2007).
The simulated $\delta Y$ is typically small ($\sim 0.03$),
however, such a smaller $\delta Y$
is consistent with  $\delta Y$ observed in 1G
stars within GCs.
We here suggest that $\delta Y$ observed in 1G and 2G stars of GCs 
is caused 
by star formation from gas polluted by massive stellar winds within
GC-forming GMCs and by
that from AGB ejecta (mixed with pristine gas), respectively.

\subsection{Discrete multiple stellar populations}

One of key observational results related to the origin of multiple
stellar populations in GCs is that
the distribution of stars
along the [Mg/Fe]-[Al/Fe] anti-correlation in NGC 2808 is not
continuous (Carretta 2014):
NGC 2808 consists of three distinct groups with
different [Mg/Fe] and [Al/Fe].
This type of discrete multiple stellar populations has been
discovered in other GCs, such as NGC 6752 and M22 
(e.g., Carretta et al. 2012; Milone et al 2013; Marino et al. 2011).
BJP17 proposed the following scenario that explains the
origin of these discrete multiple stellar populations.
First,  the first generation of stars  (1G) are formed from
a giant molecular  gas.
Then, about 30 Myr after 1G formation,
the second generation of new stars (2G)
 can be formed from AGB ejecta of 1G population.
This 2G stars can last only for [10-20] Myr,
because of the gas expulsion by the 2G's most massive
stars with $m_{\rm s}$ significantly lower than that of the 1G's most massive stars. 
After the truncation of 2G
star formation by SNe,
the third generation (3G) of stars
are then formed from AGB ejecta.
Thus, the origin of discrete multiple stellar populations is 
due to 
this cycle of star formation followed by its truncation
by SNe in BJP17.

It is assumed in BJP17 that (i) star formation  can be resumed soon after
low-mass SNe ($m=8 {\rm M}_{\odot}$) occurred
and thus (ii) the time lag between two subsequent stellar 
populations ($t_{\rm lag}$) 
is $\sim 3 \times 10^7$ yr.
However, as shown in the present study, these assumptions would not be
so reasonable
owing to $f_{\rm g, th}$. After the total removal of remaining gas from a GC,
an enough amount of gas from AGB stars needs to be accumulated so that
$f_{\rm g}$ can exceed $f_{\rm g, th}$. Therefore, $t_{\rm lag}$
can be significantly longer than $\sim 3 \times 10^7$ yr adopted in BJP17,
which means that the discreteness in the distribution 
 along the Mg-Al anti-correlation could be even more pronounced.
Furthermore, $f_{\rm g, th}$ suggests that star formation can be naturally 
truncated when $f_{\rm g}$ becomes lower than
$f_{\rm g, th}$.
BJP17 adopted an assumption that
all AGB ejecta is removed from
the GC by some unknown physical process in order to avoid everlasting star
formation.
The present study suggests that
there is no need for theories of GC formation
to adopt an ad hoc assumption of star formation truncation thanks to $f_{\rm g, th}$.

This scenario, however, has not demonstrated that gas chemically polluted
by 1G (2G) SNe cannot participate in the formation of 2G (3G) stars: [Fe/H] 
should be very similar between different generations of stars to explain
the observed very small [Fe/H] spreads ($<0.05$ dex) for normal GCs. 
In particular,
it is not so clear whether AGB ejecta used for the formation
of  2G stars in a GC can be completely
ejected from the GC so that AGB ejecta from 2G stars for 3G formation
 can have
almost the same  [Fe/H] as 1G and 2G stars. 
The present study has demonstrated that if $f_{\rm g} < f_{\rm g, th}$,
then star formation can be severely truncated even without SN feedback 
effects. Accordingly, 2G star formation is truncated when
$f_{\rm g}$ becomes lower than $f_{\rm g, th}$, and 3G star formation
cannot start until $f_{\rm g}$ becomes higher than
$f_{\rm g, th}$. This provide an alternative scenario
(without SN feedback effects)  that 
the origin of discrete MSPs is closely associated 
with  $f_{\rm g, th}$.

If $m_{\rm u}$ is less than $8 {\rm M}_{\odot}$, then only 2G (no 3G, 4G etc) can be formed
in the present AGB scenario. Accordingly, the scenario can explain the origins of
GCs with discrete  (e.g., NGC 2808) or continues abundance spreads self-consistently.
It is not clear how other GC formation scenarios can explain the discrete MSPs observed
in  some  GCs. Elmegreen (2017) proposed a new GC formation scenario in which 
stellar envelopes of high-mass 1G stars in a GC-forming GMC
are stripped and then mixed with pristine gas
to be finally converted into new stars with chemical abundances different from
those of 1G stars. He suggested that a GC-forming GMC consists of subclumps with different
self-enrichment histories, and thus that the new GC can have different discrete MSPs.
GC formation scenarios like this, in which all GC stars are formed before SNII
(i.e., within 3 Myr), can possibly explain discrete MSPs in GCs, however,
they cannot simply explain abundances spreads in $r$- and $s$-process elements
observed in some GCs: they need to invoke other physical mechanisms 
to explain such abundance spreads.
The AGB scenario can naturally explain both the discreteness of MSPs and the
abundance spreads in  $r$- and $s$-process elements in a self-consistent manner.

\subsection{Top-light IMFs  in secondary star formation ?}

The observed large fractions of 2G stars
(typically $\sim 0.7$) in GCs have been suggested to require either original GC
masses that are by a factor of $\sim 10$ larger than the present-day ones
or a particular combination of IMFs for 1G and 2G stars
(e.g., Bekki \& Norris 2006; Prantzos \& Charbonnell 2006).
The required large mass of a GC is known as the ``mass budget'' problem,
which has not been solved yet.
One of possible  ways to alleviate this mass budget problem is to assume that
the IMF for 2G stars is top-light, because the mass fractions of low-mass 2G stars
(i.e., those which can be observed in the present-day GCs) can be significantly
larger than those for a canonical IMF. 
As shown in the present study, the formation of massive gas clumps that
lead to the formation of massive stars with $m_{\rm s} \ge 8 {\rm M}_{\odot}$
can be severely suppressed in GCs. 
The combination of a top-heavy IMF for 1G stars and a top-light IMF for
2G stars can significantly reduce the required initial masses
of GCs in comparison with models with canonical IMFs
both for 1G and 2G stars (Bekki \& Norris 2006).

Such a top-light IMF can allow  secondary star formation to last longer,
because the mass (lifetime) of the most massive star in secondary star 
formation can be small (longer). As shown in our previous studies 
(B17b),  star formation from AGB ejecta can continue to be efficient,
only if feedback effects of SNII from 2G stars are  suppressed in GCs,
because such feedback effects can expel most of AGB ejecta within
GCs.  
Such a high star formation efficiency due to much less efficient
SNII feedback effects in  GC formation can increase the final mass
fraction of 2G stars.
Cabrera-Ziri et al. (2014) concluded that there is no star formation
in massive young star clusters within disk galaxies,
because no H$\beta$ and [OIII] emission from  massive OB stars was found 
(see Goudfrooij et al. 2014 for a criticism for this interpretation of the
observational results).
This observed apparent
lack of massive OB stars in young massive clusters
does not necessarily mean the lack of star formation, if the IMF is
top-light (i.e., no formation of OB stars yet formation of lower mass stars).
Therefore, a top-light IMF in secondary star formation cannot only contribute
to a possible solution of the mass budget problem but also
can provide a hint for the apparent lack of secondary star formation
in young massive star clusters.

Observations have shown  that massive OB stars can be formed from direct collisions
between molecular clouds (e.g., Fukui et al. 2015, 2017, 2018). 
Recent 3D MHD simulations of colliding molecular clouds have
demonstrated that massive molecular cores, which lead to the formation
of massive stars, can be efficiently developed in the shocked gaseous layers
induced by cloud-cloud collisions (e.g., Inoue \& Fukui 2013).  Such collisions between
molecular clouds formed from gas accumulated within GCs are highly unlikely.
Therefore, if the major formation mechanism for OB star formation is
cloud-cloud collisions, then 
OB stars are highly unlikely to be formed within GCs. 
Thus, these recent observational and theoretical studies of OB star formation through
cloud-cloud collisions also suggest top-light IMFs in secondary star formation within
GCs.


\section{Conclusions}

We have investigated whether gravitational interaction between gas and individual
stars (``stellar bombardment", SB) within a GC can influence secondary star formation in the GC using
our new hydrodynamical simulations. 
Many models with different gas mass fractions ($f_{\rm g}$), initial GC masses and sizes
($M_{\rm gc}$ and $R_{\rm gc}$, respectively), and gas disk sizes ($R_{\rm g}$)
have been investigated
so that the effects of SB on star formation can be discussed.
The main conclusions are as follows. \\

(1) Small gas clouds with $\rho_{\rm g} > 10^{10}$ atom cm$^{-3}$
corresponding to first stellar cores can be formed due to
local gravitational instability  within
gas disks without initial turbulence.
Consequently,
a significant fraction of the gas 
can be converted into new stars within a short timescale 
($< 3 \times 10^6$ yr).  This  star formation
in gas disks without initial turbulence 
is in a striking contrast with star formation processes
due to turbulent fragmentation
of self-gravitating gas clouds that have been investigated
extensively in many previous works (e.g., Klessen et al. 1998). \\

(2) SB can suppress the growth of high-density gas clumps,
if $f_{\rm g}$ is less than a threshold gas fraction 
($f_{\rm g, th} \sim 0.01$ for $M_{\rm gc} \sim 3 \times 10^5 {\rm M}_{\odot}$).
This suppression ends up with delayed and less efficient  secondary
star formation in GCs with low $f_{\rm g}$. 
Secondary star formation proceeds efficiently
with $\epsilon_{\rm sf}>0.5$ for  $f_{\rm g}>f_{\rm g, th}$.
This $f_{\rm g, th}$ is smaller for larger $M_{\rm gc}$, which implies that
secondary star formation from gas can start earlier in more massive GCs. \\

(3) GCs with larger $R_{\rm g}$ show larger $f_{\rm g, th}$ for
$M_{\rm gc} \ge 3 \times 10^5 {\rm M}_{\odot}$.
Since the original angular momentum of GCs (1G stars) can determine the sizes
of gas disks (B11), this result implies that the angular momentum
is one of key factors for star formation from the accumulated gas within GCs. \\

(4) Gas ejected from massive AGB stars can be accumulated in the 
central region of a GC earlier, because more massive AGB stars have shorter
lifetimes. The He abundances of  massive AGB stars
are predicted to be quite high
(e.g., Ventura \& D'Antona 2009; Karakas 2010).  Therefore, if 2G stars are
formed from AGB ejecta
earlier, then they can have high He abundances.  
Such 2G stars with high He abundances are likely to be formed in
massive GCs where $f_{\rm g, th}$ is lower, i.e., star formation
can start earlier. \\

(5) SB and gravitational potentials of GC can combine to  suppress
the formation of  massive and high-density gas clumps
with $m_{\rm c}\ge 8 {\rm M}_{\odot}$, which implies that
the formation of massive stars that can explode as SNII can be suppressed
in GCs. This possible top-light IMF implies that the mass budget problem
of GC formation
is less severe than was suggested by previous theoretical models. Furthermore,
such top-light IMF ensures that
secondary star formation 
within GCs can last longer owing to the lack of energetic
SNII of short-lived massive stars 
that can expel all of the remaining gas from GCs. 
It is suggested  that young stellar objects
within massive clusters with ages of several $10^7$ yr
can have the top-light IMF, if they exist in the clusters. \\

(6) The derived $f_{\rm g, th}$ suggests that gas can be kept
in the central regions of GCs without been converted into new stars
for a significantly longer time scale. This can increase the probability
that gas abundance in $r$-process elements
ejected from neutron star merging can be trapped in GCs,
because such trapping of high-speed ejecta requires high densities of
intra-cluster gas (e.g., BT17).
Therefore,  $f_{\rm g, th}$ can be closely related to  the origin of 
abundances spreads in $r$-process elements in some GCs in the 
Galaxy. \\

(7) Thus, dense stellar systems such as GCs
and stellar galactic nuclei can be a ``double edge sword'' for 
star formation within the systems.
  Deep gravitational potential well of such systems
can retain gas ejected from existing stars such as AGB stars.
Such retained gas can be used for secondary star formation within the systems.
However,
SB can suppress the mass growth of small gas clumps and thus secondary
star formation within the systems, if the gas mass fractions are less
than $f_{\rm g, th}$ for star formation. 
This $f_{\rm g, th}$ can result in
discrete epochs of star formation,  bursty nature of
secondary star formation,  retention of gas from merging of
neutron stars and possibly from SNIa and delayed SNII in 
the early evolution of GCs. \\

\section{Acknowledgment}
I (Kenji Bekki; KB) am   grateful to the referee  for  constructive and
useful comments that improved this paper.

\end{document}